\documentclass[3p,number,final,12pt,longtitle]{elsarticle}
\makeatletter\if@twocolumn\PassOptionsToPackage{switch}{lineno}\else\fi\makeatother

\usepackage{amsmath}
\usepackage{caption}
\usepackage{mfirstuc}
\usepackage{tabulary,xcolor}
\usepackage{amsfonts,amsmath,amssymb}
\usepackage[T1]{fontenc}
\makeatletter
\let\save@ps@pprintTitle\ps@pprintTitle
\def\ps@pprintTitle{%
  \let\@oddhead\@empty
  \let\@evenhead\@empty
  \let\@evenfoot\@oddfoot
}
\def\hlinewd#1{%
  \noalign{\ifnum0=`}\fi\hrule \@height #1%
  \futurelet\reserved@a\@xhline}

\AtBeginDocument{\ifNAT@numbers \biboptions{sort&compress}\fi}
\makeatother

\usepackage{ifluatex}
\ifluatex
\usepackage{fontspec}
\defaultfontfeatures{Ligatures=TeX}
\usepackage[]{unicode-math}
\unimathsetup{math-style=TeX}
\else 
\usepackage[utf8]{inputenc}
\fi 
\ifluatex\else\usepackage{stmaryrd}\fi
\usepackage{url,multirow,morefloats,floatflt,cancel,tfrupee}
\makeatletter

\AtBeginDocument{\@ifpackageloaded{textcomp}{}{\usepackage{textcomp}}}
\makeatother
\usepackage{colortbl}
\usepackage{xcolor}
\usepackage{pifont}
\usepackage[nointegrals]{wasysym}
\makeatletter

\def\mcWidth#1{\csname TY@F#1\endcsname+\tabcolsep}

\def\cAlignHack{\rightskip\@flushglue\leftskip\@flushglue\parindent\z@\parfillskip\z@skip}
\def\rAlignHack{\rightskip\z@skip\leftskip\@flushglue \parindent\z@\parfillskip\z@skip}

\@ifundefined{etal}{}{}

\usepackage{ifxetex}
\ifxetex\else\if@twocolumn\@ifpackageloaded{stfloats}{}{\usepackage{dblfloatfix}}\fi\fi

\AtBeginDocument{
\expandafter\ifx\csname eqalign\endcsname\relax
\def\eqalign#1{\null\vcenter{\def\\{\cr}\openup\jot\m@th
  \ialign{\strut$\displaystyle{##}$\hfil&$\displaystyle{{}##}$\hfil
      \crcr#1\crcr}}\,}
\fi
}

\AtBeginDocument{%
  \@ifpackageloaded{endfloat}%
   {\renewcommand\efloat@iwrite[1]{\immediate\expandafter\protected@write\csname efloat@post#1\endcsname{}}}{\newif\ifefloat@tables}%
}%

\def\BreakURLText#1{\@tfor\brk@tempa:=#1\do{\brk@tempa\hskip0pt}}
\let\lt=<
\let\gt=>
\def\processVert{\ifmmode|\else\textbar\fi}

\@ifundefined{subparagraph}{
\def\subparagraph{\@startsection{paragraph}{5}{2\parindent}{0ex plus 0.1ex minus 0.1ex}%
{0ex}{\normalfont\small\itshape}}%
}{}

\newcommand\role[1]{\unskip}
\newcommand\aucollab[1]{\unskip}
  
\@ifundefined{tsGraphicsScaleX}{\gdef\tsGraphicsScaleX{1}}{}
\@ifundefined{tsGraphicsScaleY}{\gdef\tsGraphicsScaleY{.9}}{}
\def\checkGraphicsWidth{\ifdim\Gin@nat@width>\linewidth
	\tsGraphicsScaleX\linewidth\else\Gin@nat@width\fi}

\def\checkGraphicsHeight{\ifdim\Gin@nat@height>.9\textheight
	\tsGraphicsScaleY\textheight\else\Gin@nat@height\fi}

\def\fixFloatSize#1{}
\let\ts@includegraphics\includegraphics

\def\inlinegraphic[#1]#2{{\edef\@tempa{#1}\edef\baseline@shift{\ifx\@tempa\@empty0\else#1\fi}\edef\tempZ{\the\numexpr(\numexpr(\baseline@shift*\f@size/100))}\protect\raisebox{\tempZ pt}{\ts@includegraphics{#2}}}}

\AtBeginDocument{\def\includegraphics{\@ifnextchar[{\ts@includegraphics}{\ts@includegraphics[width=\checkGraphicsWidth,height=\checkGraphicsHeight,keepaspectratio]}}}

\DeclareMathAlphabet{\mathpzc}{OT1}{pzc}{m}{it}

\def\URL#1#2{\@ifundefined{href}{#2}{\href{#1}{#2}}}

\def\UrlOrds{\do\*\do\-\do\~\do\'\do\"\do\-}%
\g@addto@macro{\UrlBreaks}{\UrlOrds}

\edef\fntEncoding{\f@encoding}

\makeatother

\newif\ifmultipleabstract\multipleabstractfalse%
\usepackage{float}
\usepackage[utf8]{inputenc}
\usepackage[english]{babel}

\usepackage{blindtext}
\usepackage[utf8]{inputenc}
\usepackage[english]{babel}
\usepackage{fancyhdr}
\usepackage{lastpage}
 
\usepackage[english]{babel}
\usepackage[utf8]{inputenc}
\usepackage{fancyhdr}

\setcitestyle{square}
\usepackage{subcaption}
\usepackage{graphicx}
\usepackage{amsmath}
\usepackage{amsmath,amssymb,amsfonts}
\usepackage{algorithmic}
\usepackage{graphicx}
\usepackage{textcomp}
\usepackage{xcolor}
\usepackage{comment}
\usepackage{mfirstuc}
\newcommand{\quotes}[1]{``#1''}
\def\BibTeX{{\rm B\kern-.05em{\sc i\kern-.025em b}\kern-.08em
    T\kern-.1667em\lower.7ex\hbox{E}\kern-.125emX}}
\usepackage{mfirstuc}
\usepackage{hyperref}
\usepackage{url}

\begin{document}


\pagenumbering{arabic}

\begin{frontmatter}
\thispagestyle{empty}
	
\title{Peering Costs and Fees}
\author[]{Ali Nikkhah\fnref{fn1}}
\author[]{Scott Jordan\fnref{fn2}}

\fntext[fn1]{Ali Nikkhah is a 5\textsuperscript{th} year Ph.D. candidate in Networked Systems at the University of California, Irvine. He obtained the B.Sc. and M.Sc. degrees in Electrical Engineering from Sharif University of Technology in 2016 and 2018, respectively. He was a data science intern at Autodesk (June 2021 - March 2022). Email: ali.nikkhah@uci.edu}
\fntext[fn2]{Scott Jordan is a Professor of Computer Science at the University of California, Irvine. He served as the Chief Technologist of the Federal Communications Commission during 2014-2016. Email: sjordan@uci.edu. Mailing address: 3214 Bren Hall, Department of Computer Science, University of California, Irvine, CA 92697-3435. Webpage: www.ics.uci.edu/~sjordan/ \\
This material is based upon work supported by the National Science Foundation under Grant No. 1812426.
}
\begin{abstract}
\thispagestyle{empty}
Internet users have suffered collateral damage in tussles over paid peering between large ISPs and large content providers. Peering between two networks may be either settlement-free or paid. Paid peering is a relationship where two networks exchange traffic with payment, which provides direct access to each other's customers without having to pay a third party to carry that traffic for them. On the other hand, two networks will agree to settlement-free peering if this arrangement is superior for both parties compared to alternative arrangements including paid peering or transit. The conventional wisdom is that two networks agree to settlement-free peering if they receive an approximately equal value from the arrangement. In order to qualify for settlement-free peering, large Internet Service Providers (ISPs) require that peers meet certain requirements. However, the academic literature has not yet shown the relationship between these settlement-free peering requirements and the value to each interconnecting network. 
 
We first consider the effect of paid peering on broadband prices. We adopt a two-sided market model in which an ISP maximizes profit by setting broadband prices and a paid peering price. We model two broadband plans: a basic plan and a premium plan for consumers with significant incremental utility from video streaming. Our result shows that paid peering fees reduce the premium plan price, and increase the video streaming price and the total price for premium tier customers who subscribe to video streaming services. ISP profit increases but video streaming profit decreases as an ISP moves from settlement-free peering to paid peering price. 
 
We next consider the effect of paid peering on consumer surplus. We simulate a regulated market in which a regulatory agency determines the maximum paid peering fee (if any) to maximize consumer surplus, an ISP sets its broadband prices to maximize profit. We find that consumer surplus is a uni-modal function of the paid peering fee. The paid peering fee that maximizes consumer surplus depends on elasticities of demand for broadband and for video streaming. However, it does not follow that settlement-free peering is always the policy that maximizes consumer surplus. The peering price depends critically on the incremental ISP cost per video streaming subscriber; at different costs, it can be negative, zero, or positive. 
 
Last, we construct a network cost model to analyze the effect of the number and location of interconnection points, routing policy, and content replication on the incremental ISP cost per video streaming subscriber. We show that the traffic-sensitive network cost decreases as the number of interconnection points increases, but with decreasing returns. Interconnecting at 6  to 8 interconnection points is rational, and requiring interconnection at more than 8 points is of little value. We show that if the content delivery network (CDN) delivers traffic to the ISP locally, then a requirement to interconnect at a minimum number of interconnection points is rational. We also show that if the CDN delivers traffic using hot potato routing, the ISP is unlikely to perceive sufficient value to offer settlement-free peering. 
 
\end{abstract}
\begin{keyword} 
Broadband\sep Regulation\sep Net Neutrality\sep Two-sided Model\sep Internet Interconnection \sep Paid Peering
\end{keyword}

\end{frontmatter}

\section{Introduction}

We focus on peering. Historically, peering was principally used by Tier 1 networks. Peering may be either paid (i.e., one interconnecting network pays the other) or settlement-free (i.e., without payment). The conventional wisdom is that two Tier 1 networks agree to settlement-free peering if and only if the two networks perceive a roughly equal exchange of value from the peering arrangement. For example, if two Tier 1 networks are both ISPs with similar numbers of customers and similar size backbones, then they may perceive a roughly equal value from the exchange of traffic with destinations in their customer cones. Large ISPs often require that peers meet certain requirements, including a specified minimum number of interconnection points, a traffic ratio less than 2:1, and symmetric routing. The conventional wisdom is that these requirements are related to the perception of roughly equal value, but the academic literature has not yet established such a relationship. 

It is no longer clear who should pay whom and how much for interconnection between Internet Service Providers (ISPs) and content providers. Large ISPs claim that large content providers are imposing a cost on ISPs by sending large amounts of traffic to their customers. ISPs claim that it is more fair that content providers pay for this cost than consumers, because then this cost will be paid only by those consumers with high usage. In contrast, large content providers (including CDNs) claim that when they interconnect with ISPs at interconnection points (IXPs) close to consumers, they are already covering the costs of carrying traffic through the core network, and that consumers are already covering the costs of carrying traffic through the ISP's access network. These disputes between large ISPs and large content providers have recurred often during the last 10 years. When not resolved, large ISPs have often refused to increase capacity at interconnection points with large content providers and transit providers, resulting in sustained congestion which has degraded users' quality of experience because of reduced throughput, increased packet loss, increased delay, and increased jitter.

In some countries, including the United States, there has been an increasing number of disputes over interconnection between large ISPs, on one side, and large content providers and transit providers, on the other side. In 2013-2014, a dispute between Comcast and Netflix over terms of interconnection went unresolved for a substantial period of time, resulting in interconnection capacity that was unable to accommodate the increasing Netflix video traffic. In 2014, Netflix and a few transit providers brought the issue to the attention of the United States Federal Communications Commission (FCC), which was writing updated net neutrality regulations. The FCC discussed the dispute in the 2015 Open Internet Order \citep{FCC}. 

Both large ISPs and large content providers agree that settlement-free peering is appropriate when both sides perceive equal value to the relationship. However, whereas large content providers assert that carrying their traffic to an interconnection point close to consumers is of value, large ISPs assert that "if the other party is only sending traffic, it is not contributing something of value to the broadband Internet access service provider".

In 2015, the FCC was concerned about the duration of unresolved interconnection disputes and about the impact of these disputes upon consumers. However, it concluded that in 2015 it was "premature to draw policy conclusions concerning new paid Internet traffic exchange arrangements between broadband Internet access service providers and [content] providers, CDNs, or backbone services." Thus, in 2015 the FCC adopted a case-by-case approach in which it would monitor interconnection arrangements, hear disputes, and ensure that ISPs are not engaging in unjust or unreasonable practices. However, in 2018, the FCC reversed itself and ended its oversight of interconnection arrangement, when it repealed most of the 2015 net neutrality regulations \citep{FCC2}. It is almost certain that the FCC will revisit the issue in the next few years.

The goal of this paper is to evaluate the effect of paid peering fees on broadband prices and consumer surplus. We then evaluate the effect of cost on paid peering. Finally, we evaluate the effect of number of interconnection points and video traffic localization by the video streaming provider on ISP direct peering cost. Our principal approach is to model the interaction between an ISP and its subscribers, and between an ISP and large content provider, as a two-sided market model. To the best of our knowledge, this is the first work to use a two-sided market model to analyze the effect of paid peering fees on broadband prices and consumer surplus.

The rest of this paper is organized as follows. Section \ref{sec:Literature} summarizes the relevant research literature. Section \ref{sec:model} proposes a model of user subscription to broadband service tiers and to video streaming. We consider a monopoly ISP that offers basic and premium tiers differentiated by bandwidth and price. To focus on the effect of peering fees, in Section \ref{sec:two-sided} we propose a two-sided model in which a monopoly ISP maximizes its profit by choosing broadband prices as well as a peering price. In Section \ref{sec:effect_prices}, we consider the effect of paid peering on broadband prices as well as ISP profit. In Section \ref{sec:consumer_surplus}, we consider the impact of paid peering on consumer surplus. In Section \ref{sec:Cd}, we show that the peering price depends critically on the incremental ISP cost per video streaming subscriber, and that at different costs it can be negative, zero, or positive. 

In Section \ref{sec:Model}, we develop a cost model in which subscribers are distributed according to census statistics and the ISP's network consists of access networks, middle mile networks, and a backbone. In Section \ref{Sec:traffic_costs}, we determine the distances on each portion of an ISP's network over which it carries traffic to and from an end user. In Section \ref{sec:CDN}, we analyze interconnection between a large content provider and an ISP.

\section{Research Literature}\label{sec:Literature}

A few papers examine the effects of interconnection agreements in the Internet backbone by using two-sided market models. 

Kim \cite{kim2020direct} is concerned with whether an ISP that is vertically integrated with a content provider may use peering fees to gain advantages over unaffiliated content providers. It proposes a two-sided market model with one monopoly ISP, one affiliated content provider, and one unaffiliated content provider. The research problem addressed in Kim \cite{kim2020direct} differs from that which we consider here. First, Kim \cite{kim2020direct} is focused on the effect of a peering price on competition between content providers, while we focus on the effect on both content providers and consumers. Second, Kim \cite{kim2020direct} adopts a game theoretic approach, while we consider both profit maximization and consumer surplus maximization. 

Laffont et al. \cite{laffont2003internet} are concerned with how interconnection fees between a pair of ISPs affect the allocation of network costs between consumers and content providers. It considers a two-sided model in which there is perfect competition between two ISPs, each of which can serve any customer or content provider. Although there are some parallels between the results of Laffont et al. \cite{laffont2003internet} and the results of our paper, the issues and models are quite different, since Laffont et al. \cite{laffont2003internet} are concerned with interconnection fees between two competitive ISPs whereas we are concerned with interconnection fees between a monopoly ISP and content providers.

Wang et al. \cite{wang2018paid} are concerned with how interconnection fees between an ISP and content providers affect ISP profit and consumer surplus. It proposes a two-sided model in which a monopoly ISP may provide content providers the choice between paid peering and settlement-free peering and in which the ISP charges consumers an amount proportional to their monthly usage. Although both Wang et al. \cite{wang2018paid} and our paper are concerned with the impact of interconnection fees on both ISP profit and consumer surplus, Wang et al. \cite{wang2018paid} are focused primarily on the ISP decision of how much capacity to allocate to paid versus settlement-free peering, whereas we are focused primarily on the ISP decision of the peering price.

In addition to these three papers that use two-sided models to examine issues relating to interconnection, there is a much larger set of papers that use two-sided models to examine issues relating to net neutrality. Most of these papers are concerned with the impact of a paid prioritization prohibition on ISP profit and consumer surplus \cite{musacchio2007network,weisman2010price,economides2012network,njoroge2014investment,tang2019regulating}. Also, there is an even larger set of papers that use two-sided models to analyze other aspects of various telecommunications markets \cite{ma2008interconnecting,wu2011revenue,wang2017optimal}.

Finally, there are also some papers that model the benefits and costs of peering between a CDN and an ISP. Chang et al. \cite{chang2006peer} propose benefit-based and cost-based frameworks for interconnection decisions by ISPs. They suggest that large ISPs choose peers based on their geographic scope and number of customers, and the traffic ratio. Agyapong and Sirbu \cite{sirbu2011economic} examined the relationship between ISPs and CDNs and proposed a model of how routing or interconnection choices might influence total costs and potential payment flows. However, neither paper considers the number or location of interconnection points, nor routing.

As a result, the academic literature provides limited insight into how to judge disputes between ISPs and content providers over interconnection. In 2014, as part of the Federal Communications Commission's net neutrality proceeding, some large content providers and some large ISPs disagreed over the appropriate requirements for settlement-free peering between content providers and ISPs. For example, Verizon asserted that \quotes{[i]f parties exchange roughly equal amounts of traffic ..., then the parties may exchange traffic on a settlement-free basis}, but that \quotes{when the traffic exchange is not roughly balanced, then the net sending party typically makes a payment in order to help compensate the net receiving party for its greater relative costs to handle the other party’s traffic} \cite{VerizonReplyComments}. In contrast, Netflix asserted that \quotes{[traffic] [r]atio-based charges no longer make economic sense since traffic ratios do not accurately reflect the value that networks derive from the exchange of traffic} \cite{NetflixComments}.

\section{A Model of User Subscription to Broadband and to Video Streaming} \label{sec:model}

Before we can analyze the effect of paid peering on broadband prices, we need a model of user subscription to broadband service tiers and to video streaming.

\subsection{Service offerings} \label{sec:services}

ISPs offer multiple tiers of broadband services, differentiated principally by download speed. ISPs typically market these broadband service tiers by recommending specific tiers to consumers who engage in specific types of online activities. For example, Comcast recommends a lower service tier to consumers who principally use their Internet connection for email and web browsing, but a higher service tier to consumers who use the Internet for video streaming. Much of the debate over paid peering concerns consumers who stream large volumes of video. Thus, we construct here a model that includes two broadband service tiers: a basic tier with a download speed intended for email, web browsing, and a limited amount of video streaming; and a premium tier (at a higher price) with a download speed intended for a substantial amount of video streaming. Although most often ISPs offer more than two tiers, the majority of customers subscribe to a subset of two tiers, and this two-tier model is sufficient to separately evaluate the effect of paid peering prices on consumers who utilize video streaming and on consumers who don’t. 

Specifically, we model a single monopoly ISP that offers a basic tier at a monthly price $P^b$ and a premium tier at a monthly price $P^b+P^p$. We consider $N$ consumers, each of whom may subscribe to the basic tier, the premium tier, or neither. We denote user $i$'s utility per month from subscription to the basic tier by $b_i$, and user $i$'s utility per month from subscription to the premium tier by $b_i+p_i$. We presume that a consumer who gains significant utility from video streaming subscribes to the premium tier. 

To analyze the effect of paid peering prices on broadband prices, we focus on the aggregate of all video streaming providers that directly interconnect with the ISP and that may pay (or be paid) a fee for peering with the ISP. We model the aggregate of all plans offered by these video streaming providers, but to keep the model tractable we consider a single price of $P^v$ per month for the aggregate. We denote user $i$'s utility per month from subscription to video streaming providers by $v_i$. Consumer $i$'s utility from all other content is included in $b_i+p_i$.

Consumers differ in the utilities they place on broadband service tiers and on video streaming. We assume that the number of consumers $N$ is large. 

\subsection{Demand functions} \label{sec:demand}

Each consumer thus has four choices
\begin{equation} \label{options}
X_i \triangleq \left\{\begin{array}{lr}
n, & \text{do not subscribe} \\
b, & \text{subscribe to the basic tier} \\
p, & \text{subscribe to the premium tier but not to a video streaming provider} \\
v, & \text{subscribe to the premium tier and to a video streaming provider}.
\end{array} \right.
\end{equation}

Consumer $i$'s consumer surplus, defined as utility minus cost, under each choice is thus
\begin{equation}\label{CSi}
CS_i(X_i;b_i,p_i,v_i) = \left\{ {\begin{array}{lr}
0, & X_i=n \\
b_i - P^b, & X_i=b \\
b_i + p_i - P^b - P^p, & X_i=p \\
b_i + p_i + v_i - P^b - P^p - P^v, & X_i=v.
\end{array}} \right.
\end{equation}

Each consumer is assumed to maximize consumer surplus. Thus, consumer $i$ adopts the choice
\begin{equation} \label{choice}
X_i^*(b_i,p_i,v_i) \triangleq \arg \max_{X_i} CS_i(X_i;b_i,p_i,v_i),
\end{equation}
and earns a corresponding consumer surplus $CS_i^* \triangleq CS_i(X_i^*)$. 

Each of the $N$ consumers makes an individual choice per (\ref{choice}). The consumers who choose to subscribe to the basic tier are those whose utility $b_i$ from subscription to the basic tier exceeds its monthly price $P^b$, whose incremental utility $p_i$ from subscription to the premium tier without subscribing to a video streaming provider falls below the incremental monthly price $P^p$, and whose incremental utility $p_i+v_i$ from subscription to the premium tier and to video streaming falls below the corresponding incremental monthly price $P^p+P^v$. Thus, the demand\footnote{Since we model a finite number $N$ of consumers whose utilities are given by a joint probability density function, this equation, and other similar equations below, give the \textit{average demand}. However, for simplicity of presentation, we use the term \textit{demand}.} for the basic tier is given by $N^b(P^b,P^p,P^v)$.

Similarly, the consumers who choose to subscribe to the premium tier but not to video streaming are those whose utility $b_i+p_i$ from subscription to the premium tier exceeds its monthly price $P^b+P^p$, whose incremental utility $p_i$ from subscription to the premium tier without subscribing to video streaming exceeds the incremental monthly price $P^p$, and whose incremental utility $v_i$ from subscription to video falls below the incremental monthly price $P^v$. Thus, the number of consumers who subscribe to the premium tier but who do not subscribe to video streaming is given by $N^p(P^b,P^p,P^v)$.

Finally, the consumers who choose to subscribe to both the premium tier and video streaming are those whose utility $b_i+p_i+v_i$ from subscription to both services exceeds the combined cost $P^b+P^p+P^v$, whose incremental utility $p_i+v_i$ from subscription to only the basic tier exceeds the corresponding incremental price $P^p+P^v$, and whose incremental utility $v_i$ from subscription to video streaming falls exceeds the incremental monthly price $P^v$. Thus, the demand for video streaming is given by $N^v(P^b,P^p,P^v)$.

The demand for the premium tier is $N^p+N^v$, the sum of the demands for the premium tier without and with a subscription to the streaming video provider.

\subsection{Consumer surplus} \label{sec:surplus}

The aggregate consumer surplus will be an important quantity to consider in our deliberations below. Given a set of prices, the aggregate consumer surplus of subscribers to the basic tier is $CS^b(P^b,P^p,P^v)$. Similarly, the aggregate consumer surplus of consumers who subscribe to the premium tier but not to video streaming is $CS^p(P^b,P^p,P^v)$ and the aggregate consumer surplus of consumers who subscribe to both the premium tier and video steaming is $CS^v(P^b,P^p,P^v)$.

The aggregate consumer surplus over all consumers is defined as
\begin{equation} \label{CS}
CS(P^b,P^p,P^v) \triangleq CS^b(P^b,P^p,P^v) + CS^p(P^b,P^p,P^v) + CS^v(P^b,P^p,P^v) 
\end{equation}

\subsection{Profits} \label{sec:profits}

We assume that the ISP incurs a monthly marginal cost $C^b$ per basic tier subscriber. The ISP marginal profit per basic tier subscriber is thus $P^b-C^b$. We assume that the ISP incurs a monthly marginal cost $C^b+C^p$ per premium tier subscriber who does not also subscribe to video streaming. The ISP marginal profit per such broadband service tier subscriber is thus $P^b+P^p-C^b-C^p$. We also assume that the capacity is fixed in our model.

The marginal cost to an ISP associated with video streaming is at the core of the debate over paid peering, and thus we must be careful in its formulation. Here, we have assumed that only premium tier subscribers engage in a substantial amount of video streaming, consistent with ISP marketing of their service tiers. We have further divided premium tier subscribers according to whether they also subscribe to video streaming services that have direct interconnection with the ISP. 

For generality, we thus associate an ISP monthly marginal cost $C^b+C^P+C^d$ per video streaming subscriber, where the $d$ denotes direct interconnection. The incremental ISP cost $C^d$ per video streaming subscriber may be negative, zero, or positive. It is critical to note that this incremental cost is not that of the interconnection point itself between the ISP and each video streaming provider, as the cost of the interconnection point itself is negligible. However, there are several variables that may affect the incremental ISP cost per video streaming subscriber. First, video streaming subscribers receive substantially more traffic per month than premium tier subscribers who don't subscribe to video streaming. Second, when a content provider switches from indirect interconnection through a transit provider to an ISP to direct interconnection with the ISP, the location of the interconnection point may change. This change in the location of the interconnection point may result in either shorter or longer paths on the ISP's network from the interconnection point to the subscriber, and thus either a lower or higher incremental ISP cost per video streaming subscriber. 

We also consider a peering price of $P^d$ per video streaming subscriber for direct interconnection between the ISP and video streaming providers. This price may be positive if the ISP charges video streaming providers for direct interconnection, negative if the video streaming providers charge the ISP for direct interconnection, or zero if the peering is settlement-free. 

The ISP marginal profit per video streaming subscriber is $P^b+P^p+P^d-C^b-C^p-C^d$. The total ISP profit (excluding fixed costs)\footnote{Throughout the paper, ISP profit excludes fixed costs.} is thus
\begin{equation} \label{isp_profit}
\pi^{ISP}(P^b,P^p,P^d,P^v) = (P^b-C^b)N^b + (P^b+P^p-C^b-C^p)N^p + (P^b+P^p+P^d-C^b-C^p-C^d)N^v.
\end{equation}

We assume that the video streaming providers incur a monthly marginal cost $C^v$ per subscriber. The aggregate video streaming provider marginal profit per subscriber is thus $P^v-C^v-P^d$, and their total profit (excluding fixed costs)\footnote{Throughout the paper, aggregate video streaming profit excludes fixed costs.} is
\begin{equation} \label{cp_profit}
\pi^{VSP}(P^b,P^p,P^d,P^v) = (P^v-C^v-P^d)N^v.
\end{equation}

\section{A Two-Sided Model for ISP Profit Maximization} \label{sec:two-sided}

The previous section presented a model for consumer demand for broadband and video streaming, resulting in the demand functions, the corresponding aggregate consumer surplus, and the corresponding ISP and video streaming provider profits (\ref{isp_profit}-\ref{cp_profit}). In this section, we formulate a two-sided model of how the prices are determined.

\subsection{Analytical model} \label{sec:two-sided analytical}

There are a number of options for modeling how the broadband service tier prices ($P^b$ and $P^p$), the video streaming price ($P^v$), and the peering price ($P^d$) are determined. 

Throughout the paper, we presume that the ISP has no significant competition for broadband service at acceptable speeds within the footprint of its service territory. Thus, we assume that the ISP determines its broadband service tier prices ($P^b$ and $P^p$) to maximize its profit.

A key question, critical to this analysis, is how the peering price ($P^d$) is determined. Once a subscriber chooses an ISP, the ISP has a monopoly on the transport of traffic within the ISP's access network that the customer resides in. In contrast, there may be a competitive market for the transport of Internet traffic across core networks. In this section, we assume that the location of direct interconnection between the ISP and each video streaming provider is close enough to the consumers so that all of the transport from the interconnection point to the consumers falls within the ISP's access network. Correspondingly, we assume that the ISP has the market power to determine the peering price ($P^d$) and that it sets this price to maximize its profit.

The ISP thus chooses the broadband service tier prices ($P^b$ and $P^p$) and the peering price ($P^d$) to maximize its profit, namely
\begin{equation} \label{isp}
(P_{ISP}^b,P_{ISP}^p,P_{ISP}^d) = \arg \max_{(P^b,P^v,P^d)} \pi^{ISP}(P^b,P^p,P^d,P^v)
\end{equation}

In contrast, we assume that the market determines the aggregate video streaming price, excluding paid peering fees, when there is no regulation of prices. We denote the aggregate video streaming price, excluding paid peering fees, by $P^{v}_0$. We presume that an ISP charging peering prices would likely charge them to both directly interconnected content providers and directly interconnected transit providers. We further presume that transit providers would pass peering prices through to their customers. As a consequence, we foresee that peering prices would be paid by all large video service providers selling to the ISP's customers. An open question is whether the video streaming providers can pass through any peering price ($P^d$) to their customers by adding it to their video streaming prices. We denote the pass-through rate of the peering fee by $0 < \alpha \leq 1$:
\begin{equation}\label{vsp}
P^v(P^d) = {P^{v}_0}+\alpha P^d
\end{equation}


Equations (\ref{isp}-\ref{vsp}) set up a two-sided model in which the ISP earns revenue from both its customers and video service providers (if $P^d>0$). The combination of the two equations captures the inter-dependencies between the ISP, the video services providers, and the consumers. The ISP-determined peering price ($P^d$), along with the pass-through rate ($\alpha$), leads to an aggregate video streaming price ($P^v)$. The ISP-determined broadband service tier prices ($P^b$ and $P^p$), along with the aggregate video streaming price ($P^v)$, lead to demands for each broadband service tier ($N^b$ and $N^p+N^v$) and for video streaming services ($N^v$). These demands in turn affect how the ISP sets each of the prices.

Since the aggregate video service price ($P^v$) is solely determined by (\ref{vsp}), we can represent the ISP's profit as a function of three variables rather than four:
\begin{equation} \label{isp2}
(P_{ISP}^b,P_{ISP}^p,P_{ISP}^d) = \arg \max_{(P^b,P^v,P^d)} \pi^{ISP}(P^b,P^p,P^d,{P^{v}_0}+\alpha P^d)
\end{equation}

\subsection{Numerical parameters} \label{sec:Parameters}

This two-sided model is somewhat amenable to closed-form analysis. However, we find it useful to also examine the model under a set of realistically chosen parameters. We set out those parameters in this subsection.

The joint probability density function of user utilities for the basic tier, the premium tier, and video streaming is represented by $f_{B,P,V}(b,p,v)$. For numerical evaluation, we assume that each utility is independent and has a Normal distribution: $B \sim \mathcal{N}(\mu_b,\,\sigma_b^{2}), P \sim \mathcal{N}(\mu_p,\,\sigma_p^{2}), V \sim \mathcal{N}(\mu_v,\,\sigma_v^{2})$. We need to determine numerical values for the means and variances.

The ISP incurs a monthly marginal cost of $C^b$ per subscriber, a monthly marginal cost of $C^p$ per premium tier subscriber, and an incremental ISP cost $C^d$ per video streaming subscriber. We need to determine numerical values for these three costs.

Unfortunately, direct information about user utilities and ISP costs is scarce. Instead, we choose numerical values for user utilities and ISP costs indirectly using available information about demand and prices in the United States.

There are several sets of publicly available statistics about broadband prices and subscriptions \citep{Pew,WSJ}. While the set of statistics differ, they show that roughly 75\% of households in the United States subscribe to fixed broadband service. Hence, we wish to choose numerical values for user utilities and ISP costs so that, at the ISP profit-maximizing prices, $(N^b+N^p+N^v)/N = 0.75$. For each ISP, the statistics show that subscribers predominately choose among two service tiers, which we map to the basic and premium tiers modeled above, with roughly 2/3 of subscribers choosing the premium tier. Hence, we wish to choose numerical values for user utilities and ISP costs so that, at the ISP profit-maximizing prices, $N^b/N = (0.75)(1/3) = 0.25$ and $(N^p+N^v)/N = (0.75)(2/3) = 0.50$. Moreover, the statistics also reveal that the price of the lower of the two popular tiers is roughly \$50 per month, and the price of higher of the two popular tiers is roughly \$70 per month. Hence, we wish to choose numerical values for user utilities and ISP costs so that the ISP profit-maximizing prices are $P^b=\$50.00$ and $P^p=\$20.00$.

According to \cite{Deloitte}, Americans subscribe to an average of three paid video streaming providers. There are also several sets of publicly available statistics about video streaming prices and subscriptions \citep{Leichtman,Parks}. While the set of statistics differ, they show that roughly $50\%$ of households in the United States that subscribe to fixed broadband service also subscribe to at least two video streaming services. Hence, we wish to choose numerical values for user utilities and ISP costs so that, at the ISP profit-maximizing prices, $N^v/N = (0.75)(0.5) = 0.375$.

There is even less information about the variance of user utilities, or correspondingly about the elasticity of demand. We choose $\sigma_b = \mu_b/4$, $\sigma_p = \mu_p/4$, and $\sigma_v = \mu_v/4$, which results in reasonably wide distributions.\footnote{The results below are not very sensitive to these choices.}

From these statistics, we can generate targets for the ISP profit-maximizing broadband prices $P^b$ and $P^p$, and for the demands $N^b$, $N^p$, and $N^v$ at these prices. We cannot, however, use these statistics to generate a target for the ISP profit-maximizing peering fee $P^d$, since information about these fees is scarce. Instead, we estimate the incremental ISP cost $C^d$ per video streaming subscriber. Unfortunately, we know very little about ISP network costs. We obtain a target of $C^d=\$3.00$ per month per video subscriber. That said, later in this paper, we will consider a wide range of values of $C^d$.

This now gives us six target values ($P^b$, $P^p$, $N^b$, $N^p$, $N^v$, and $C^d$) to determine the six desired parameters ($\mu_b$, $\mu_p$, $\mu_v$, $C^b$, $C^p$, and $P^d$). We can use the three equations for demand and the ISP profit maximization equation (\ref{isp2}) to determine these six desired parameters. The result is: $\mu_b \approx \$56.12$, $\mu_p \approx \$18.91$, $\mu_v \approx \$27.67$, $C^b \approx \$16.50$, $C^p \approx \$19.00$, and $P^d \approx \$4.59$. In addition, we assume the aggregate video streaming price ${P^{v}_0}=\$21.58$, based on the aggregate price of the three most popular video streaming services.\footnote{The sum of the advertised prices of the lowest price plans for Netflix, Hulu/Disney+, and HBO Max is \$26.17\citep{Netflix,Hulu,Disney,HBO}. From this sum, we subtract the peering fee $P^d = \$4.59$.} In addition, although we consider any pass-through rate of the paid peering fee ($0 < \alpha \leq 1$), in the numerical results below we use $\alpha=1$.

We use these parameters in the remainder of the paper except as noted.

\section{The Effect of Paid Peering on Prices} \label{sec:effect_prices}

We now consider the effect of paid peering on broadband prices. ISPs assert that paid peering revenue is offset by lower broadband prices, and that ISP profits remain unchanged. Content providers assert that peering prices do not result in lower broadband prices, but simply increase ISP profits. The goal is this section is to evaluate these assertions.

With an understanding of how the ISP sets the prices $(P_{ISP}^b,P_{ISP}^p,P_{ISP}^d)$, we can now evaluate the impact of the peering price $P^d$ upon the broadband prices $P^b$ and $P^p$.

As we did in the previous section, we assume that the video streaming price $P^v$ is set by (\ref{vsp}). However, whereas in (\ref{isp2}) the ISP sets the peering price $P^d$ to maximize profit, in this section we make the peering price $P^d$ an independent variable so that we can judge its impact on other prices.

Given a specified peering price $P^d_{reg}$, the ISP is assumed to choose the tier prices $P^b$ and $P^p$ so as to maximize profit, namely
\begin{equation} \label{reg1}
(P_{reg}^b,P_{reg}^p) = \arg \max_{(P^b,P^p)} \pi^{ISP}(P^b,P^p,P^d_{reg},{P^{v}_0}+\alpha P^d_{reg}).
\end{equation}

The ISP chosen prices $(P_{reg}^b,P_{reg}^p)$ are a function of the independently set price $P^d_{reg}$. The video streaming price $P^v$ is also a function of $P^d_{reg}$. 

\begin{figure}[!tbp]
  \centering
  \begin{minipage}[t]{0.49\textwidth}
    \includegraphics[width=\textwidth]{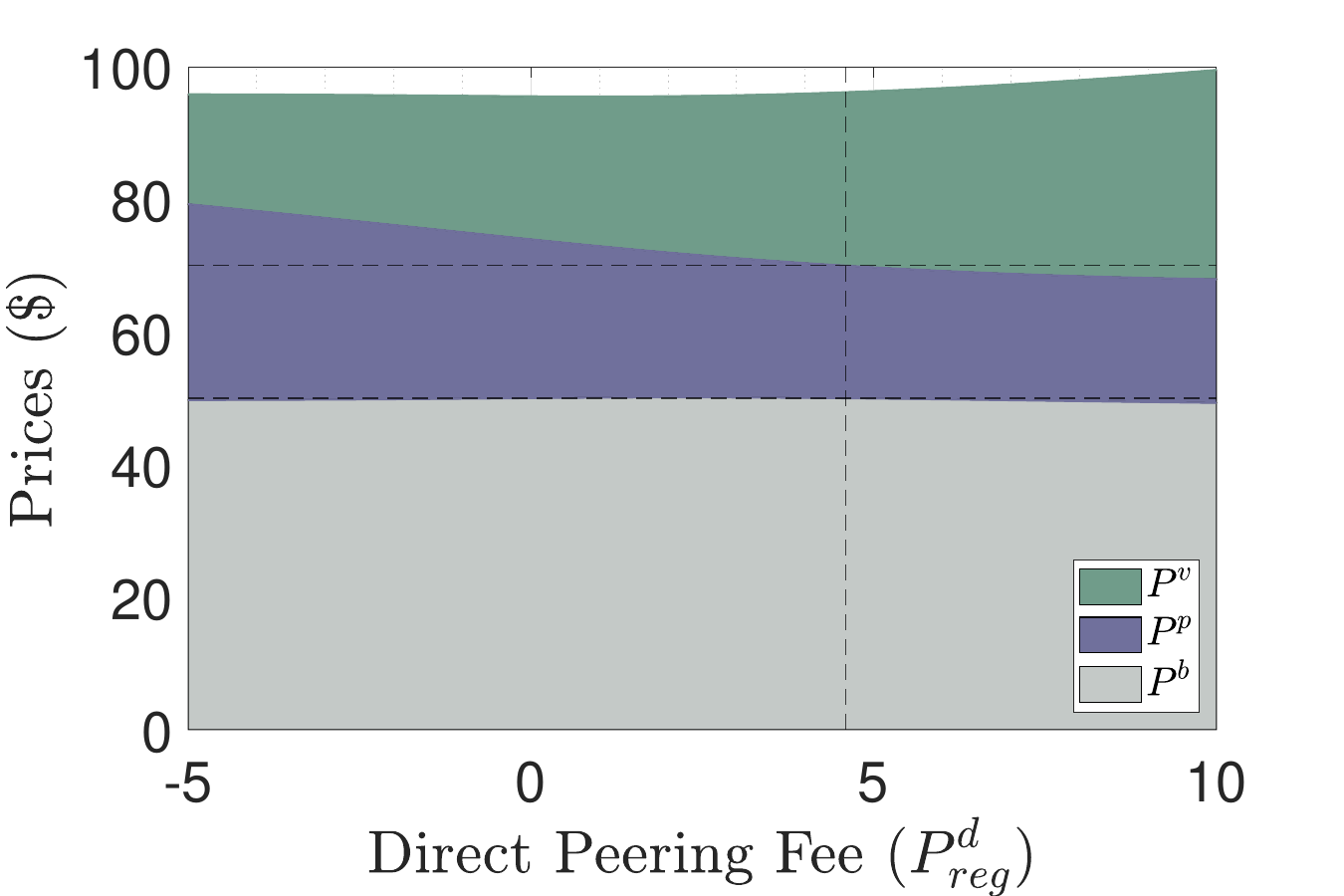}
    \caption{Effect of Peering Fee on Broadband Prices and the Aggregate Video Streaming Price}
    \label{fig:price}
  \end{minipage}
  \hfill
  \begin{minipage}[t]{0.49\textwidth}
    \includegraphics[width=\textwidth]{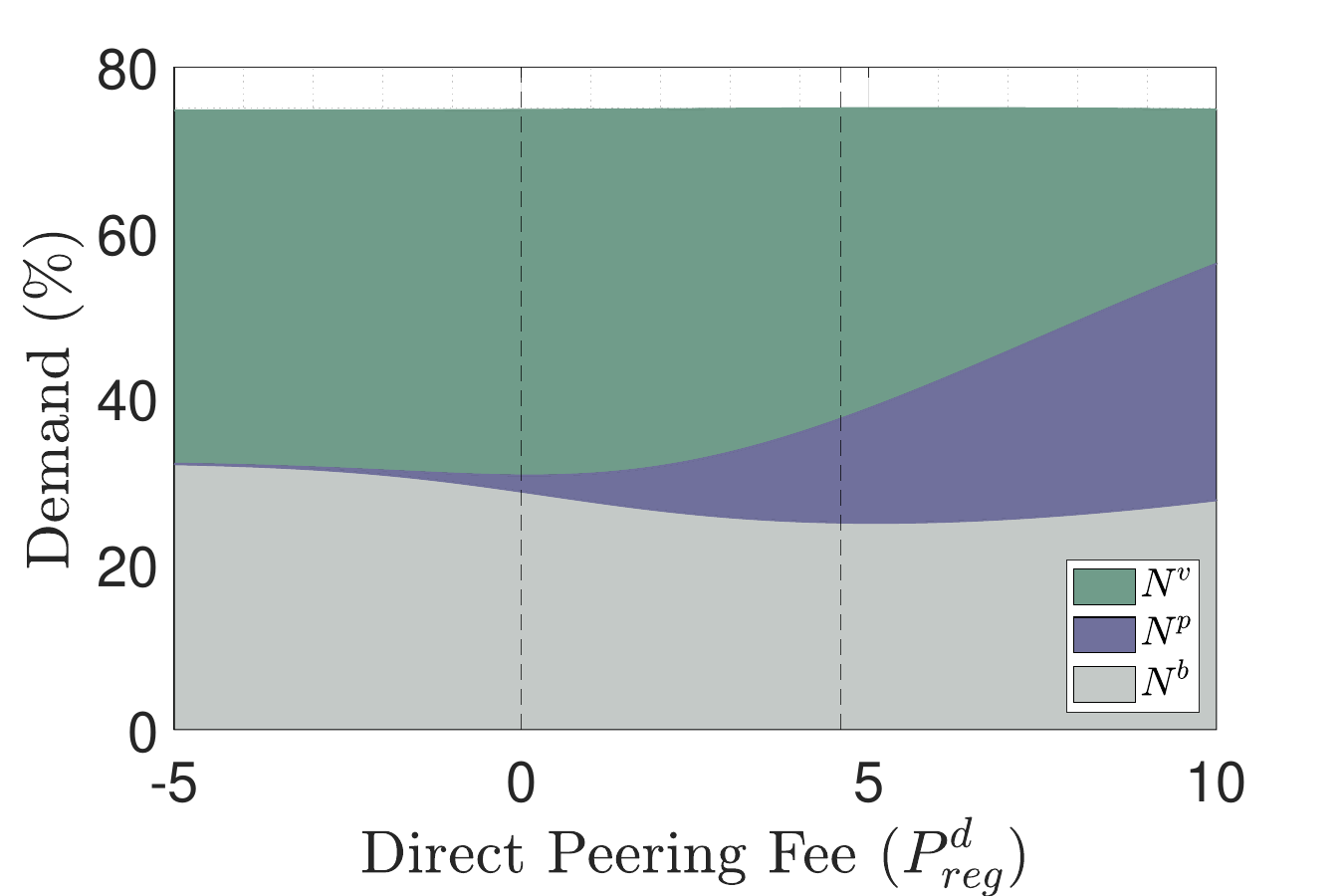}
    \caption{Effect of Peering Fee on Demand (Percentage to Total)}
    \label{fig:demands}
  \end{minipage}
\end{figure}

Figure \ref{fig:price} shows the prices of both broadband tiers and the aggregate video streaming price as a function of the independently chosen peering fee $P^d_{reg}$. 

We initially compare prices and profits in the case in which the ISP chooses the peering price to maximize profit ($P^d=\$4.59$) to the case in which settlement-free peering is used (i.e., $P^d=\$0$). We start at the profit-maximizing peering price $P^d=\$4.59$ and consider a small decrease. If the ISP did not change the prices for the broadband tiers (which it will), then a small decrease in the peering price would result in a small decrease in demand for the basic tier, a small decrease in demand for the premium tier without video streaming, and a small increase in demand for the premium tier with video streaming. 

However, the ISP now has the motivation to modify the broadband tier prices. The decrease in the peering price results in a decrease in the aggregate price of video streaming. As a consequence, the ISP will recoup most of the decreased peering price by  increasing the incremental price for the premium tier $P^p$. It does not, however, change the basic tier price $P^b$ by much at all, since increasing the premium tier price results in some users downgrading to the basic tier, which more than offsets those who would otherwise upgrade from the basic tier to the premium tier to take advantage of lower video streaming prices. The signs of these trade-offs remain the same in the entire range from $P^d=\$4.59$ to $P^d=\$0$.

Figure \ref{fig:demands} shows the corresponding demands for each broadband tier and for video streaming. Again, we start at the profit-maximizing peering price $P^d=\$4.59$ and consider a small decrease. The ISP's increase in the premium tier price drives some consumers who subscribe to the premium tier but not to video streaming to downgrade to the basic tier. However, the total price for the premium tier and video streaming, $P^b+P^p+P^v$, decreases, and thus some consumers who subscribe to the premium tier but not to video streaming now choose to start subscribing to video streaming.

\begin{figure}[!tbp]
  \centering
  \begin{minipage}[t]{0.49\textwidth}
    \includegraphics[width=\textwidth]{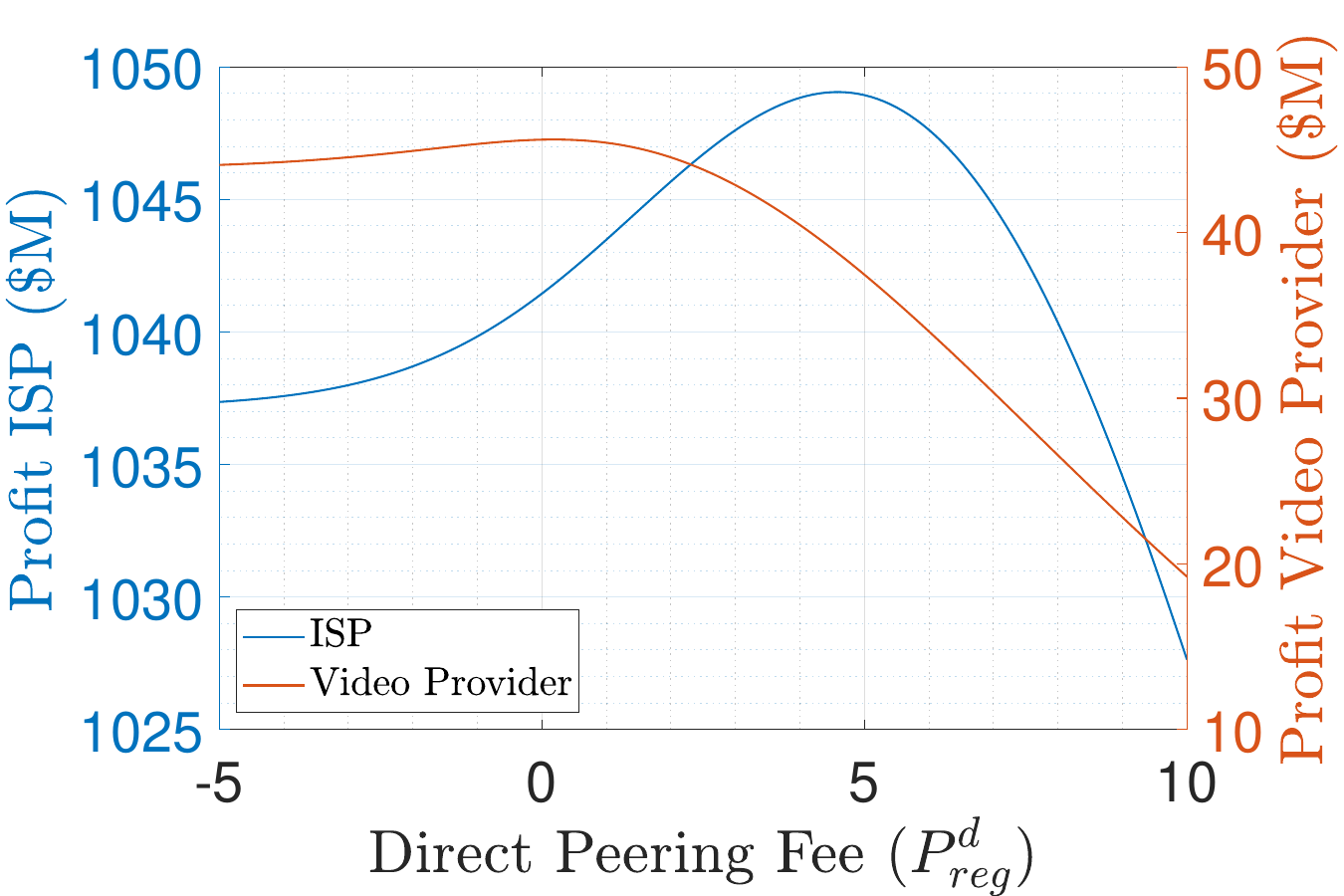}
    \caption{Effect of Peering Fee on Profit of ISP and Video Streaming Provider}
    \label{fig:profit_isp-cp}
  \end{minipage}
\end{figure}

Figure \ref{fig:profit_isp-cp} shows the corresponding ISP profit and aggregate video streaming provider profit. Again, we start at the profit-maximizing peering price $P^d=\$4.59$ and consider a small decrease. The ISP's profit from the video streaming subscribers increases because the demand $N^v$ increases and the price per subscriber $P^b+P^p$ increases. The ISP's profit from premium tier subscribers without video streaming decreases, because the demand $N^p$ decreases more than the price $P^b+P^p$ increases. Finally, the ISP's profit from basic tier subscribers increases, because the demand $N^b$ increases while the price $P^b$ remains virtually unchanged.

We can now evaluate the stakeholder claims about the effect of paid peering on broadband prices and ISP profits. Recall that ISPs assert that paid peering revenue is offset by lower broadband prices, whereas content providers assert that peering prices do not result in lower broadband prices. We find that the basic tier price $P^b$ is almost the same in the case in which the ISP chooses the peering price to maximize profit ($P^d=\$4.59$) as in the case in which settlement-free peering is used ($P^d=\$0$). We also find that the premium tier price $P^b+P^p$ decreases by \$3.98 (from \$73.98 to \$70.00) if we change from settlement-free peering ($P^d=\$0$) to paid peering ($P^d=\$4.59$), but the aggregate video streaming price increases by \$4.60 (from \$21.59 to \$26.19). Thus, to the extent that ISPs assert that paid peering reduces the price of the basic tier, we disagree. Paid peering should be expected to reduce the price of the premium tier, but this reduction in broadband price is more than offset by an increase in video streaming prices.

Recall that ISPs assert that their profits are unaffected by peering fees, whereas content providers assert that peering fees increase ISP profits. We find that the ISP profit increases by 0.8\% if we change from settlement-free peering ($P^d=\$0$) to paid peering ($P^d=\$4.59$). However, the larger effect is on aggregate video streaming profit, which decreases by 18\%.

\section{The Effect of Paid Peering on Aggregate Consumer Surplus} \label{sec:consumer_surplus}

In this section, we turn to the impact of paid peering on consumer surplus. ISPs assert that paid peering fees increase aggregate consumer surplus because they eliminate an inherent subsidy of consumers with high video streaming use by consumers without such use. Content providers assert that paid peering fees decrease aggregate consumer surplus because they are passed onto consumers through higher video streaming prices without a corresponding reduction in broadband prices.

A portion of these assertions was addressed in the previous section. We now know that when an ISP sets peering prices so as to maximize profit, it sets those prices to be positive. Compared to settlement-free peering, positive peering prices result in reduced premium tier prices. Directly connected video streaming providers increase their prices to compensate. However, the ISP only passes onto its customers a portion of the paid peering revenue. 

However, this leaves unanswered the question of the impact on aggregate consumer surplus. It also leaves unanswered the question of what value of peering price maximizes aggregate consumer surplus. We attempt to answer those questions now.

We consider the peering price $P^d$ to be an independent variable set by a regulator. The aggregate consumer surplus $CS(P_{CS_{reg}}^b,P_{CS_{reg}}^p,P^v)$ is a function of $P^d$. The regulator is presumed to set the peering price $P^d$ so that it maximizes aggregate consumer surplus:
\begin{equation} \label{CSreg}
\begin{array}{c}
(P_{CS_{reg}}^b,P_{CS_{reg}}^p) = \arg \max_{(P^b,P^p)} \pi^{ISP}(P^b,P^p,P_{CS_{reg}}^d,{P^{v}_0}+\alpha P_{CS_{reg}}^d) \\
P_{CS_{reg}}^d = \arg \max_{P^d} CS(P_{CS_{reg}}^b(P^d),P_{CS_{reg}}^p(P^d),P^d,{P^{v}_0}+\alpha P^d).
\end{array}
\end{equation}
Equation (\ref{CSreg}) determines the resulting aggregate consumer surplus maximizing value of the peering price $P^d$, as well as the resulting broadband prices $P^b$ and $P^p$ and video streaming price $P^v$.  

\begin{figure}[!tbp]
  \centering
    \begin{minipage}[t]{0.49\textwidth}
    \includegraphics[width=\textwidth]{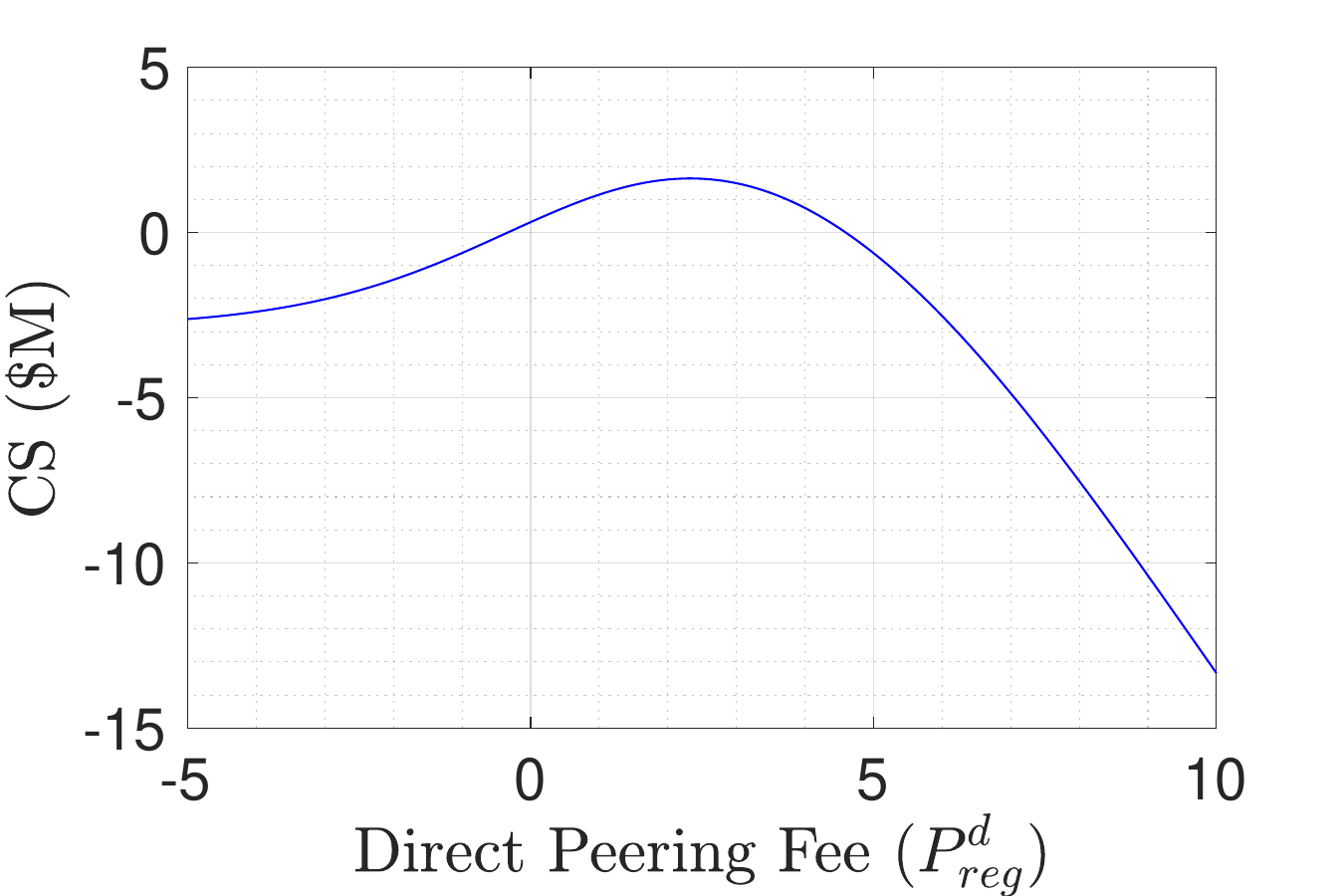}
    \caption{Effect of Peering Fee on Incremental Consumer Surplus}
    \label{fig:cs-welfare}
  \end{minipage}
    \hfill
 \begin{minipage}[t]{0.49\textwidth}
    \includegraphics[width=\textwidth]{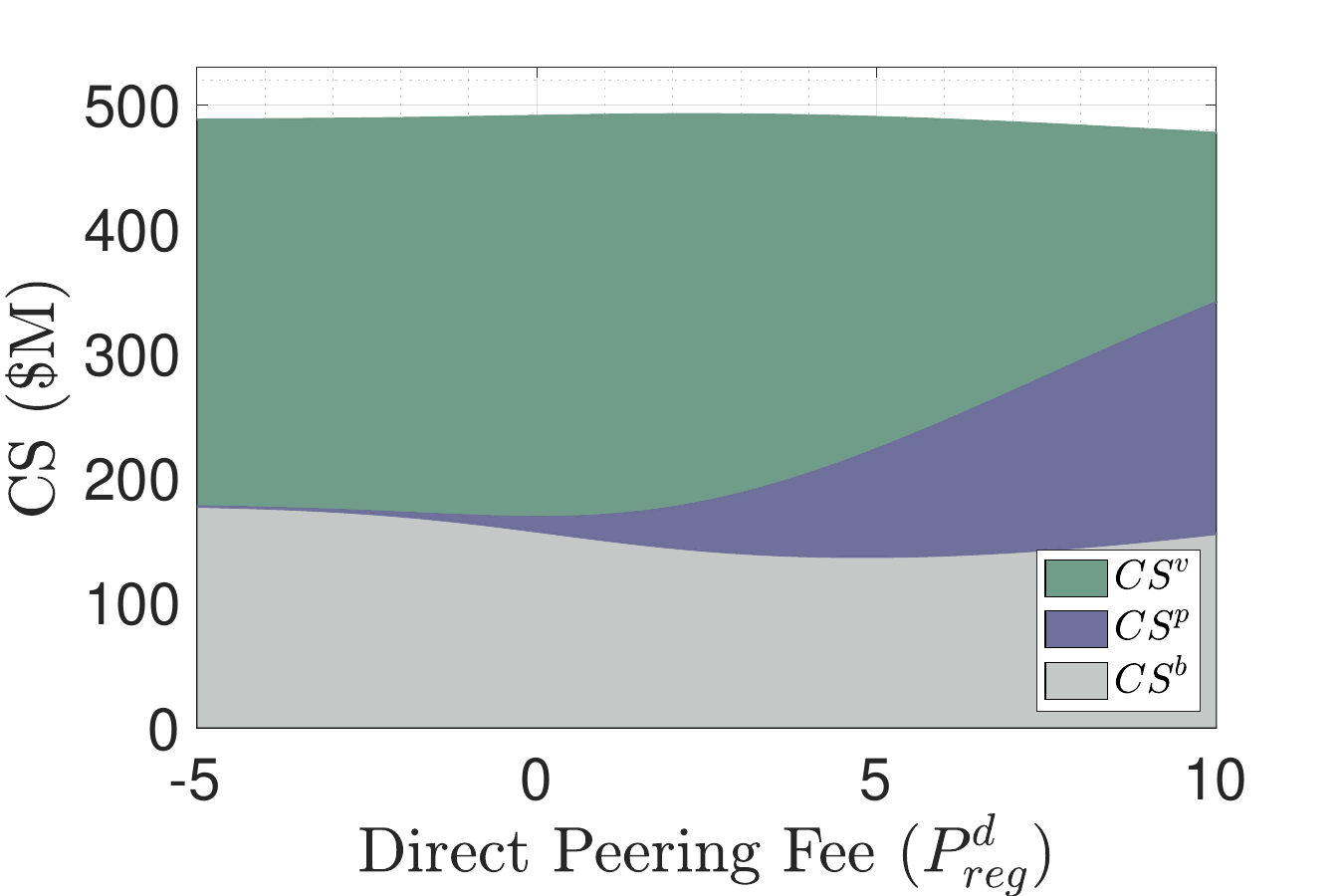}
    \caption{Effect of Peering Fee on Consumer Surplus with Different Services}
    \label{fig:surplus_tier}
  \end{minipage}
\end{figure}

Figure \ref{fig:cs-welfare} shows the incremental consumer surplus as a function of the regulator chosen peering price $P^d$. The incremental consumer surplus is defined as the difference between the aggregate consumer surplus at the regulator chosen peering price $P^d$ and at the peering price that maximizes ISP profit ($P^d_{ISP}$). 

Aggregate consumer surplus is a uni-modal function of the peering price. We find that the peering price that maximizes consumer surplus is $P^d_{CS_{reg}}=\$2.34$. This is substantially less than the peering price that maximizes ISP profit ($P^d_{ISP}$=\$4.59). At peering prices lower than \$2.34, aggregate consumer surplus decreases principally because the premium tier price is too high, and this decreases the surplus of premium tier subscribers. At peering prices higher than \$2.34, aggregate consumer surplus decreases principally because the price of video streaming is too high, and this decreases the surplus of video streaming subscribers. 

To understand why, we need to revisit the impact of the peering price on broadband tier prices and demand, and how these changes in price and demand affect aggregate consumer surplus. We compare prices and demands in the case in which the ISP chooses the peering price to maximize profit ($P^d_{ISP}=\$4.59$) to the case in which the regulator chooses the peering price to maximize aggregate consumer surplus ($P^d_{CS_{reg}}=\$2.34$). 

As we discussed in the previous section, a reduction in the peering price below that which maximizes ISP profit results in lower aggregate video streaming prices and increased premium tier prices. However, the amount of the increase in the premium tier price is less than the amount of the decrease in the aggregate video streaming price. Thus, the price of the premium tier with video streaming ($P^b+P^p+P^v$) decreases. These changes in prices cause some premium tier subscribers without video streaming to downgrade to the basic tier, and some to start subscribing to video streaming.

These changes in prices and demand affect aggregate consumer surplus. Figure \ref{fig:surplus_tier} shows the aggregate consumer surplus of all subscribers to the basic tier, to the premium tier without video streaming, and to the premium tier with video streaming. A reduction in the peering price below that which maximizes ISP profit results in increased demand for the basic tier, but with basic tier prices virtually unchanged. The result is that the aggregate consumer surplus of basic tier subscribers increases. A reduction in the peering price also results in increased premium tier prices and decreased demand for the premium tier without video streaming. The result is that the aggregate consumer surplus of premium tier subscribers without video streaming decreases. Finally, a reduction in the peering price results in decreased prices of the premium tier with video streaming and increased demand. The result is that the aggregate consumer surplus of premium tier subscribers with video streaming increases. The aggregate consumer surplus is the sum of these three. As the peering price decreases from the price that maximizes ISP profit ($P^d_{ISP}$=\$4.59) to the price that maximizes consumer surplus ($P^d_{CS_{reg}}=\$2.34$), the increase in the aggregate consumer surplus of basic tier subscribers and premium tier subscribers with video streaming dominates the decrease in the aggregate consumer surplus of premium tier subscribers without video streaming. However, at peering prices below the price that maximizes consumer surplus ($P^d_{CS_{reg}}=\$2.34$), the opposite is true.

We can now evaluate the stakeholder claims about the effect of paid peering on consumer surplus. Recall that ISPs assert that paid peering fees increase aggregate consumer surplus whereas content providers assert that they decrease aggregate consumer surplus. The peering price that maximizes aggregate consumer surplus is below the price that maximizes ISP profit. Using our numerical parameters, we found that the peering price that maximizes aggregate consumer surplus is $P^d_{CS_{reg}}=\$2.34$, whereas if unregulated the ISP would choose $P^d_{ISP}=\$4.59$. Furthermore, we found that aggregate consumer surplus is \$1.65M higher at the peering price that maximizes aggregate consumer surplus than at the peering price that maximizes ISP profit. However, we also found that when the incremental ISP cost per video streaming subscriber is $C^d=\$3.00$, aggregate consumer surplus is \$1.33M higher at the peering price that maximizes aggregate consumer surplus than at settlement-free peering ($P^d=\$0$). Thus, neither settlement-free peering nor paid peering with an ISP-determined price maximizes consumer surplus.

\section{The Effect of the Incremental ISP Cost $C^d$ Per Video Streaming Subscriber} \label{sec:Cd}

The peering price that maximizes aggregate consumer surplus depends critically on the incremental ISP cost $C^d$ per video streaming subscriber. Without knowledge of this cost, we cannot say whether the peering price that maximizes aggregate consumer surplus is negative, zero, or positive. In this section, we consider how the incremental ISP cost $C^d$ per video streaming subscriber affects the results in this paper. For each value of $C^d$, we determine the numerical parameters ($\mu_b$, $\mu_p$, $\mu_v$, $C^b$, $C^p$, and $P^d$) using the method discussed in Section \ref{sec:Parameters}. This analysis is thus a study of the impact of the unknown value of $C^d$, given fixed values for the observed known parameters.

\begin{figure}[!tbp]
  \centering
  \begin{minipage}[t]{0.49\textwidth}
    \includegraphics[width=\textwidth]{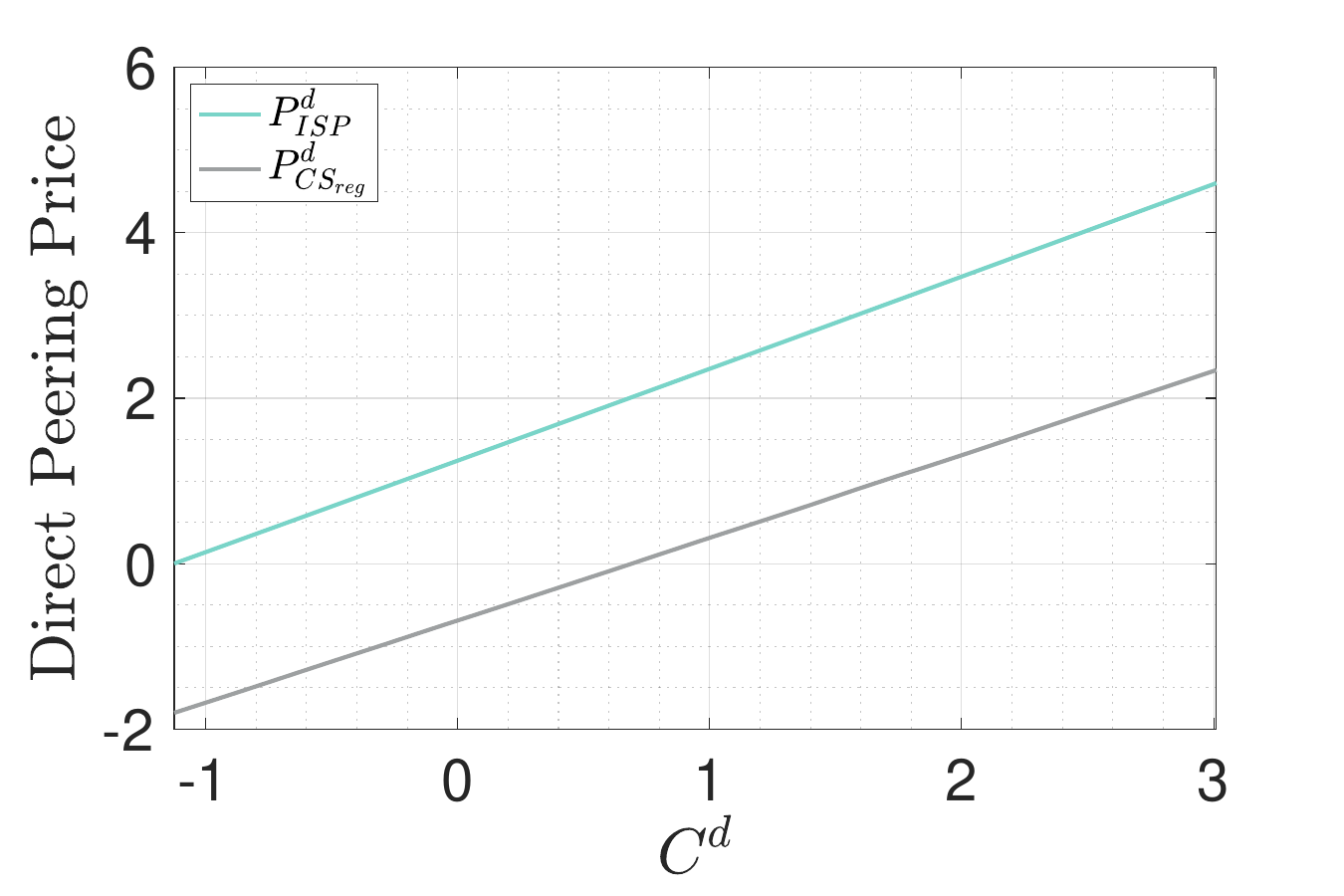}
    \caption{Effect of the Incremental ISP Cost Per Video Streaming Subscriber on the Peering Price}
    \label{Pd_vs_Cd}
  \end{minipage}
  \hfill
  \begin{minipage}[t]{0.49\textwidth}
    \includegraphics[width=\textwidth]{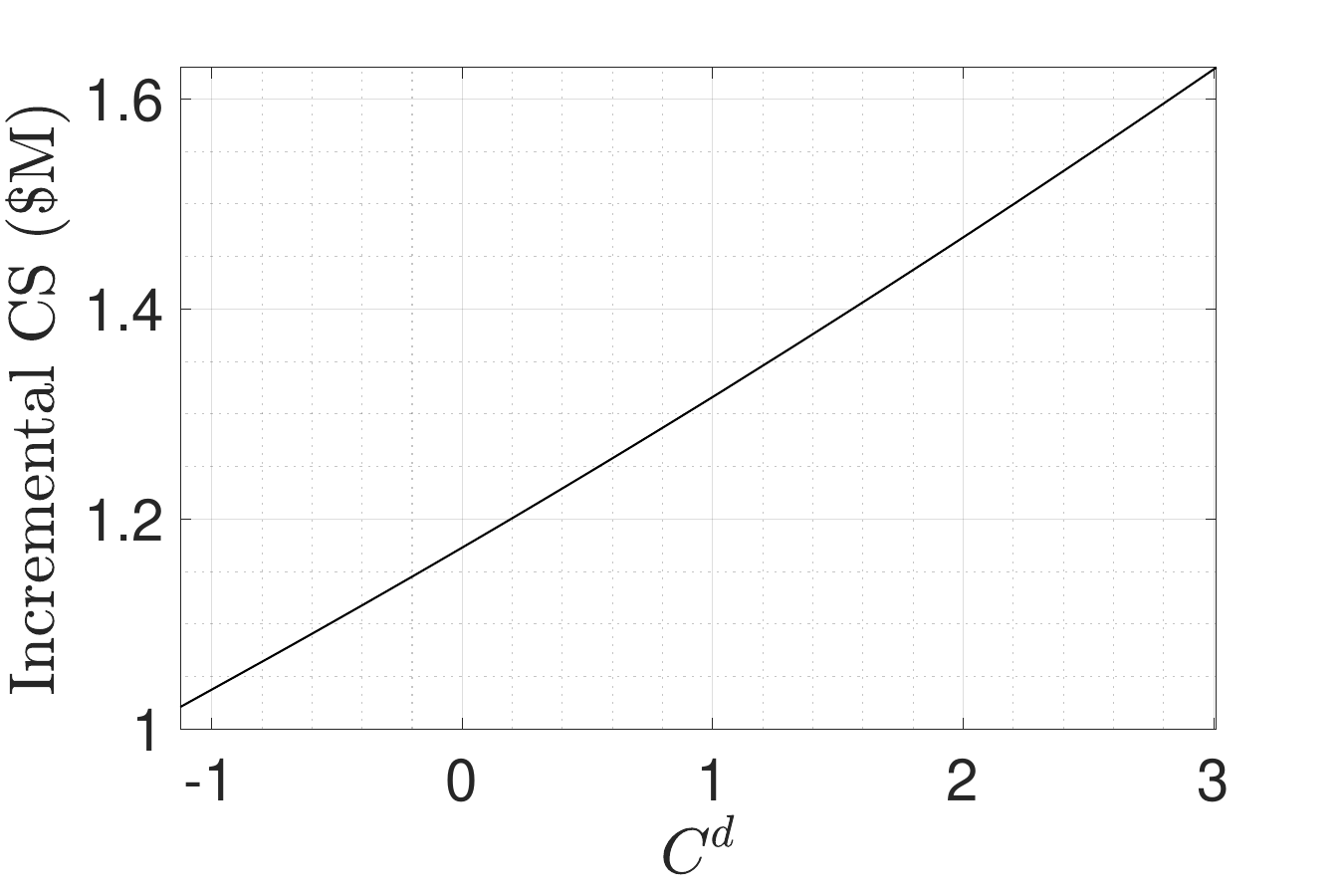}
    \caption{Effect of the Incremental ISP Cost Per Video Streaming Subscriber on the Incremental Consumer Surplus}
    \label{fig:CS_Cd}
  \end{minipage}
\end{figure}

Figure \ref{Pd_vs_Cd} shows the peering prices that maximize ISP profit and aggregate consumer surplus as a function of the incremental ISP cost $C^d$ per video streaming subscriber.Thus, not only does $C^d$ direct affect the peering prices, it also indirectly affects all prices and demands. The peering price that maximizes aggregate consumer surplus, $P^d_{CS_{reg}}$, increases nearly linearly, from -\$1.80 to \$2.34 as $C^d$ increases from -\$1.12 to \$3.00. Notably, it is positive when $C^d > \$0.68$, but negative at lower values of $C^d$. Recall that the incremental ISP cost $C^d$ per video streaming subscriber depends on both the incremental Internet usage of video streaming subscribers over non-subscribers and the length of the path on the ISP's network. As video content providers interconnect with the ISP closer to consumers, the incremental ISP cost $C^d$ per video streaming subscriber decreases, and may be negative if the interconnection point is close enough to the consumer. In contrast, if the interconnection point is far from the consumer, then the incremental Internet usage may dominate and $C^d$ may be positive.

The peering price that maximizes ISP profit, $P^d_{ISP}$, increases nearly linearly with the incremental ISP cost $C^d$ per video streaming subscriber, from \$0.00 to \$4.59 as $C^d$ increases from -\$1.12 to \$3.00. Notably, the incremental ISP profit $P^d_{ISP} - C^d$ per video streaming subscriber remains positive at all values above $C^d=-\$1.12$, and indeed increases with higher values of $C^d$.

The effect on consumers is qualitatively similar, but different in magnitude. When $C^d=\$3.00$, premium tier subscribers without video streaming would pay \$70.00 at the ISP chosen peering price ($P^d=\$4.59$) but \$71.69 if the regulator sets the peering price to maximize consumer surplus ($P^d=\$2.34$), and premium tier subscribers with video streaming would pay \$96.19 at the ISP chosen peering price but \$95.61 at the regulator chosen peering price. Thus, regulation of the peering price results in premium tier subscribers without video streaming paying \$1.69 more and in premium tier subscribers with video streaming paying \$0.58 less; however the regulated peering price also increases demand for video streaming from 37.5\% to 42.6\%. 

When $C^d=-\$1.12$, premium tier subscribers without video streaming would pay \$70.00 at the ISP chosen peering price but \$71.37 at the regulator chosen peering price, and premium tier subscribers with video streaming would pay \$91.59 at the ISP chosen peering price but \$91.15 at the regulator chosen peering price. Thus, regulation of the peering price results in premium tier subscribers without video streaming paying \$1.37 more and in premium tier subscribers with video streaming paying \$0.44 less; however the regulated peering price also increases demand for video streaming from 37.5\% to 42.3\%. 

Finally, we revisit our evaluation of stakeholder claims about broadband prices, ISP profit, and consumer surplus, under different values of the incremental ISP cost $C^d$ per video streaming subscriber. If $C^d=\$3.00$, we found that paid peering should be expected to reduce the price of the premium tier, but this reduction in broadband price is more than offset by an increase in video streaming prices. At lower values of $C^d$, paid peering still should be expected to reduce the price of the premium tier, but less so. Similarly, neither the change in ISP profit nor the change in video streaming profit is very sensitive to $C^d$. 

If $C^d=\$3.00$, we found that aggregate consumer surplus is \$1.65M higher at the peering price that maximizes aggregate consumer surplus than at the peering price that maximizes ISP profit, but that aggregate consumer surplus is also \$1.33M higher at the peering price that maximizes aggregate consumer surplus than at settlement-free peering ($P^d=\$0$). Figure \ref{fig:CS_Cd} shows the incremental consumer surplus, which is the difference between the aggregate consumer surplus at ISP-chosen peering price and that at the peering price that maximizes consumer surplus, for various values of $C^d$. We observe that the incremental consumer surplus is significant at all values of $C^d$, rising from \$1.02M to \$1.63M as $C^d$ increases from -\$1.12 to \$3.00. 

The incremental ISP cost $C^d$ per video streaming subscriber, however, does have a large impact on the optimal peering price. The peering price that maximizes consumer surplus is strongly correlated with $C^d$. At values of $C^d > \$0.68$, settlement-free peering is too aggressive. and the regulator should limit the peering price to at least \$2.00 less than the ISP-chosen peering price. At negative values of $C^d$, settlement-free peering is too timid, and the ISP should pay content providers for paid peering at locations so close to the consumers. At small positive values of $C^d$ ($0 < C^d < \$0.68$), the ISP bears a cost, but the peering price that maximizes consumer surplus is negative; we turn to this issue next.

\section{Cost Model}\label{sec:Model}

In this section, we develop a cost model that will enable our analysis of effect of number of interconnection points and traffic localization by video streaming provider on incremental ISP cost $C^d$. Thus, our model focuses on the characteristics that we believe are most critical to this analysis, and abstracts other less critical characteristics.

Our model is designed to reflect key characteristics of the United States. An ISP is assumed to serve the contiguous United States. We consider the geographic locations of the largest interconnection points (IXPs) in the United States. The network is partitioned into a backbone network, middle mile networks, and access networks. Middle mile and access networks are modeled based on U.S. counties. The density of the population is drawn from U.S. census statistics. Traffic matrices are built using these assumptions.

Subsection \ref{sec:topology} introduces the topology of an ISP's U.S. network. Subsection \ref{sec:traffic} develops the traffic matrices over this network. 

\subsection{Topology} \label{sec:topology}

The topology of an ISP's U.S. network consists of a model of the ISP's service territory, the location of IXPs, and a model of segments of the network.

\subsubsection{Service Territory} 

While most ISPs do not offer residential broadband Internet access service over the entire contiguous United States, we see little in their settlement-free peering policies that are specific to their service territory, other than that a subset of the IXPs at which they peer are concentrated near their service territory. Thus we focus on a single ISP whose service territory covers the contiguous United States. 

The ISP's service region is modeled as the contiguous United States using geographical data of its boundaries.

\subsubsection{Location of IXPs}

We focus on the interconnection between the ISP and video streaming providers. We use the geographic locations of the $M=12$ largest IXPs in the United States, listed in Table \ref{tab2} \cite{IXP4,IXP5,IXP6}. We note that the largest ISPs in the United States (Comcast, Charter, AT\&T, and Verizon) each interconnect at a minimum of 9 of these 12 IXPs, although a smaller ISP (Cox) interconnects at fewer IXPs; see Table \ref{tab2}.

\begin{table*}[!t]
\renewcommand{\arraystretch}{1.3}
\caption{The largest IXPs at which some ISPs interconnect}
\label{tab2}
\centering
\includegraphics[width=\textwidth]{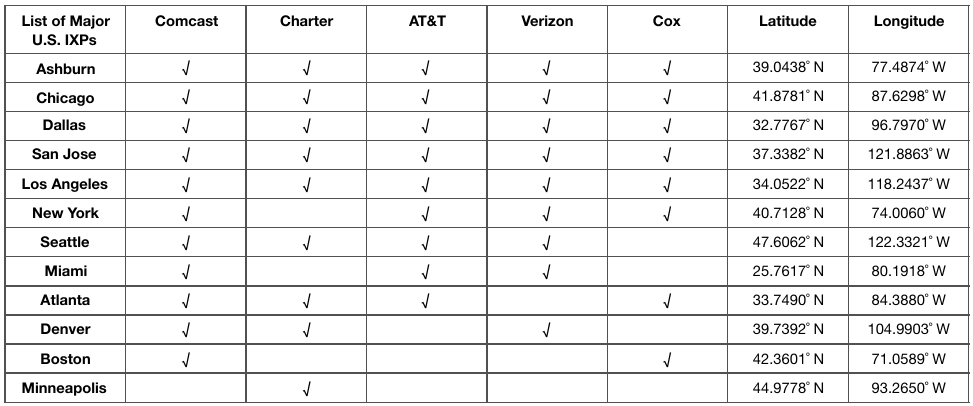}
\end{table*}

\subsubsection{Middle Mile Networks and Access Networks}

We model the ISP's network as partitioned into a single backbone network, multiple middle mile networks, and multiple access networks. We model each access network as spanning a single U.S. county. While we recognize that the geographical sizes and topologies of access networks differ widely, this assumption will not significantly affect the results in this paper, since differences in network costs between various forms of peering depend more critically on the number of interconnection points than on the topologies of access networks.

The locations of the geographical center of access networks are assigned to be the longitudes and latitudes of the center of each county in the contiguous United States \cite{Lat_Long}. A middle mile link is assumed to run from the geographical center of each access network to the closest IXP.

We consider an ISP and the video streaming provider that agree to interconnect at $N$ IXPs. Each geographical region consists of the union of access networks for which the closest IXP at which the ISP and the video streaming provider agree to peer.

\begin{figure}[!t]
\centering
\includegraphics[width=4in]{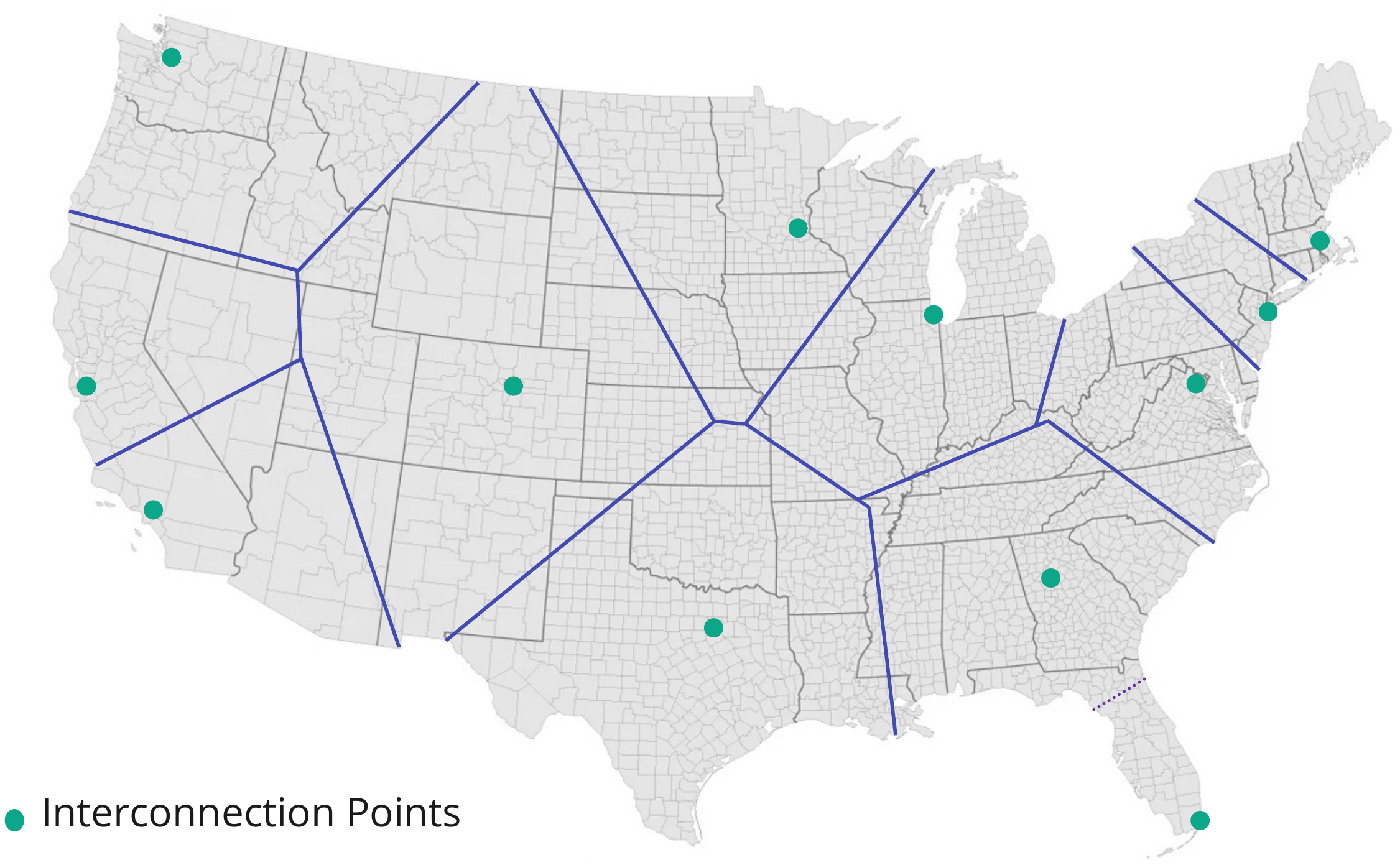}
\caption{Topology of an ISP's network}
\label{fig:1}
\end{figure}

Figure \ref{fig:1} roughly illustrates these regions when the ISP and the video streaming provider agree to interconnect all 12 IXPs.\footnote{In the figure, the partition of the regions is only roughly illustrated. More precisely, they should follow county boundaries.} 

\subsection{Traffic Matrices} \label{sec:traffic}

We now turn to modeling the traffic matrices over the ISP's network. A traffic matrix represents the volume of traffic between all pairs of sources and destinations.

\subsubsection{Distribution of Sources and End Users}\label{sub_sub_dist}

The locations of end users of the ISP are represented by a probability distribution over the ISP's service territory. We decompose this distribution into (a) a distribution of the number of end users in each access network and (b) for each access network, the distribution of end users within the access network.

We denote the probability that an end user resides within access network $j$ by $P(j)$. We assume that end users are distributed across access networks according to the population of the county associated with the access network. We denote the population of the county associated with access network $j$ by $p_j$, and we denote by $p = \sum_j p(j)$ the population of the contiguous United States. We assign these values using U.S. census data \cite{Census}. It follows that $P(j) = p_j/p$. We further assume that end users are uniformly distributed within each access network.

We focus here on downstream traffic that originates outside the ISP's network and terminates at an end user on the ISP's network. Denote the source's location by $S$ and the end user by $U$. The source $S$ may be at an IXP at which the content provider has a server. We consider the interconnection between the ISP and a content provider.

We assume that the distribution of the source $S$ is identical to the distribution of end users $U$, which is jointly given by $\{P(j)\}$ and the uniform distribution of end users within each access network. We assume that the source $S$ and the end user $U$ are independent.

In the analysis below, we distinguish between several points along traffic routes. We continue to focus on downstream traffic that originates outside the ISP's network and terminates at an end user on the ISP's network. Along the route from the source $S$ to the end user $U$, denote the location of the IXP at which downstream traffic enters the ISP's network by $IXP_{down}^{p}$ and the location of the IXP closest to the end user by $IXP^u$.

\section{Traffic-sensitive Costs} \label{Sec:traffic_costs}

Although we know that an ISP's traffic-sensitive cost is a complicated function of the topology of the network, to make the analysis tractable, we abstract the network geographically into three non-overlapping sections: backbone, middle mile, and access. We define the backbone network as the set of links between IXPs. We define the middle mile networks as the set of links between the geographical center of each access network and the closest IXP. We define the access networks as the set of links that connect the middle mile networks to end users.

In this section, we first determine the distances on each portion of its network that an ISP carries traffic from a source to an end user. We next calculate the average distance using the traffic matrices above. Finally, we model the traffic-sensitive cost associated with carrying the traffic over these average distances.

\subsection{Distances}\label{sub:distance}

For downstream traffic, the distance from $S$ to $U$ on the ISP's backbone network is a function of the location of the IXP at which downstream traffic enters the ISP's network ($IXP^p_{down}$) and the location of the IXP closest to the end user ($IXP^u$). Denote the distance on the ISP's backbone network between these two IXPs by $D^b(IXP^p_{down},IXP^u)$, the distance between $IXP^p_{down}$ and $IXP^u$.

The distance from $S$ to $U$ on the ISP's middle mile network is a function of the location of the IXP closest to the end user ($IXP^u$) and the location of the access network on which $U$ resides. Denote the distance on the ISP's middle mile network between these two locations by $D^m(IXP^u,U)$, the distance between the IXP closest to the end user ($IXP^u$) and the location of the geographical center of access network on which $U$ resides.

The distance from $S$ to $U$ on the ISP's access network is a function of the location of the end user. Denote the distance on the ISP's access network by $D^a(U)$, the distance between the location of the end user ($U$) and the geographical center of access network on which $U$ resides. The distance can be determined by the location of $U$ within the access network.

All distances in the our model are great-circle distances between the corresponding points on a sphere, and are calculated using the Haversine formula.

\subsection{Average Distances}\label{sub:av_distance}

An ISP's traffic-sensitive cost depends on the average distance of traffic on each segment of its network. As discussed above, the distances on each section of the ISP's network depend on the joint distribution of $(IXP^{p},IXP^u,U)$.

We continue to focus on downstream traffic that originates outside the ISP's network and terminates at an end user on the ISP's network. The distance on the ISP's backbone network is a function of $(IXP_{down}^{p},IXP^u)$. 

When hot potato routing is used, the IXP at which downstream traffic enters the ISP's network ($IXP_{down}^{p}$) is independent of the end user and thus independent of the IXP closest to the end user ($IXP^u$). The IXP at which downstream traffic enters the ISP's network depends on the localization policy and the IXPs at which they agree to interconnect. Consider an ISP and an content provider that agree to interconnect at $N$ IXPs. The average distance on the ISP's backbone network ($ED_{down}^{b,hot}$) is the average distance between the IXPs at which downstream traffic enter the ISP's network ($IXP_{down}^{p}$) and the IXPs closest to the end user ($IXP^u$). The probability distribution of $IXP^u$ is determined by the population of each corresponding region.

The distance on the ISP's middle mile network is a function of $(IXP^u,U)$. It is independent of the routing policy, and the average distance ($ED^m$) is the average distance between the IXPs closest to the end user and the center of the access network on which the end user resides.

The distance on the ISP's access network is a function of $U$. It is also independent of the routing policy. Since end users are uniformly distributed within each access network, but not between access networks, the average distance ($ED^a$) is the weighted average of average distances within each access network.

\subsection{Cost}\label{sub_cost}

The ISP incurs a traffic-sensitive cost for carrying traffic over the average distances calculated in the previous subsection. We consider here only traffic-sensitive costs, because non-traffic-sensitive costs do not vary with routing policies or the number of interconnection points.\footnote{There is a small cost for each interconnection point; however, this cost is relatively small compared to transportation costs.}

Traffic-sensitive costs are a function of both distance and traffic volume. We assume here that traffic-sensitive costs are linearly proportional to the average distance over which the traffic is carried on each portion of the ISP's network, see e.g., \cite{valancius2011many}. We also assume that traffic-sensitive costs are linearly proportional to the average volume of traffic that an ISP carries on each portion of its network. Although the cost might be an increasing concave function of traffic volume (or a piecewise constant function), the linear model will suffice for our analysis.

We model the cost per unit distance and per unit volume differently on the backbone network, the middle mile networks, and the access networks. Denote the cost per unit distance and per unit volume in the backbone network by $c^b$, the cost per unit distance and per unit volume in the middle mile networks by $c^m$, and the cost per unit distance and per unit volume in the access network by $c^a$. Denote the volume of traffic by $V$. The ISP's traffic-sensitive cost is thus $V \left({c^b} ED^b + {c^m} ED^m + {c^a} ED^a \right)$.

Given a fixed source-destination traffic matrix, the average distance across the ISP's access networks is constant. In addition, the average distance across the ISP's middle mile networks is constant, once we fix $M=12$, since the IXPs at which the parties agree to peer do not affect the middle mile. The variable portion of the ISP’s traffic-sensitive cost is thus only $C = c^b ED^b V$

Below we consider the effect on the variable traffic-sensitive cost ($C$) of changes in the number of IXPs at which peering occurs and localization policies, for constant $c^b$. (In the remainder of the paper, we use the term cost to refer to the variable traffic-sensitive cost.) We will find that changes in the number of IXPs at which peering occurs and routing policies affect $ED^b$.

\section{Peering between a content provider and an ISP} \label{sec:CDN}

There have been an increasing number of disputes over interconnection between large ISPs, on one side, and large content providers and transit providers, on the other side. In 2013-2014, a dispute between Comcast and Netflix over terms of interconnection went unresolved for a substantial period of time, resulting in interconnection capacity that was unable to accommodate the increasing Netflix video traffic. In 2014, Netflix and a few transit providers brought the issue to the attention of the United States Federal Communications Commission (FCC), which was writing updated net neutrality regulations. The FCC discussed the dispute in the 2015 Open Internet Order \citep{FCC}. 

The FCC first summarized the arguments of large content providers and transit providers. It noted that "[content] providers argue that they are covering the costs of carrying [their] traffic through the network, bringing it to the gateway of the Internet access service". Large content providers and transit providers argued that they should be entitled to settlement-free peering if the interconnection point is sufficiently close to consumers. The lack of willingness of large ISPs to offer settlement-free peering with large content providers, and to augment the capacity of existing interconnection points with transit providers with which they had settlement-free peering agreements, had led to the impasse. The FCC noted that "[s]ome [content] and transit providers assert that large [ISPs] are creating artificial congestion by refusing to upgrade interconnection capacity ... for settlement-free peers or CDNs, thus forcing [content] providers and CDNs to agree to paid peering arrangements." 

The FCC then summarized the arguments of large ISPs. It noted that "large broadband Internet access service providers assert that [content] providers such as Netflix are imposing a cost on broadband Internet access service providers who must constantly upgrade infrastructure to keep up with the demand". The large ISPs explained that the network upgrades include adding capacity in the middle mile and access networks. The FCC noted that the large ISPs asserted that if they absorb these costs, then the ISPs would recoup these costs by increasing the prices for all subscribers, and that the large ISPs argued that "this is unfair to subscribers who do not use the services, like Netflix, that are driving the need for additional capacity". 

Both large ISPs and large content providers agree that settlement-free peering is appropriate when both sides perceive equal value to the relationship. However, whereas large content providers assert that carrying their traffic to an interconnection point close to consumers is of value, large ISPs assert that "if the other party is only sending traffic, it is not contributing something of value to the broadband Internet access service provider".

In 2015, the FCC was concerned about the duration of unresolved interconnection disputes and about the impact of these disputes upon consumers. However, it concluded that in 2015 it was "premature to draw policy conclusions concerning new paid Internet traffic exchange arrangements between broadband Internet access service providers and [content] providers, CDNs, or backbone services." Thus, in 2015 the FCC adopted a case-by-case approach in which it would monitor interconnection arrangements, hear disputes, and ensure that ISPs are not engaging in unjust or unreasonable practices. However, in 2018, the FCC reversed itself and ended its oversight of interconnection arrangements, when it repealed most of the 2015 net neutrality regulations \citep{FCC2}. It is almost certain that the FCC will revisit the issue in the next few years.

Large ISPs do not generally have different settlement-free peering policies for content providers than for ISPs and transit providers. In addition, they have often asserted that content providers should meet the same requirements on the number of interconnection points and traffic ratio to qualify for settlement-free peering. However, It is not clear the degree to which the settlement-free peering requirements should apply to the peering between the ISP and the content provider. We will show that if a content provider delivers traffic to the ISP locally, then a requirement to interconnect at a minimum number of interconnection points is rational, but a limit on the traffic ratio is not rational. We will also show that if a content provider does not deliver traffic locally, the ISP is unlikely to perceive sufficient value to offer settlement-free peering. 

In this section, we consider a content provider that hosts a content server at each IXP at which
it agrees to peer with an ISP, but that replicates only a portion of this content and delivers only that portion to the ISP locally.

The ISP network topology remains the same as was presented in Section \ref{sec:topology}, and the distribution of the location of end users remains the same as was presented in Section \ref{sub_sub_dist}. However, the location of the content is no longer the same as in previous sections. We assume that, within each access network, a proportion $x$ of requests is served by the content server located at the IXP closest to the end user at which the content provider and the ISP agree to peer. We also assume that, within each access network, the remaining proportion $1-x$ of requests is served by a content server that is independent of the location of the end user, and that the distribution of the location of this content server is identical to the distribution of end users. We further assume that the content provider uses hot potato routing for non-locally delivered content.

The ISP's cost is:
\begin{equation}
\begin{aligned}
C_{cp}^{partial}
=c^b V_{down} \left(x ED^{b,cold}_{down}+(1-x) ED^{b,hot}_{down} \right)
\end{aligned}
\end{equation}

\begin{figure}[!t]
\centering
\includegraphics[width=0.49\textwidth]{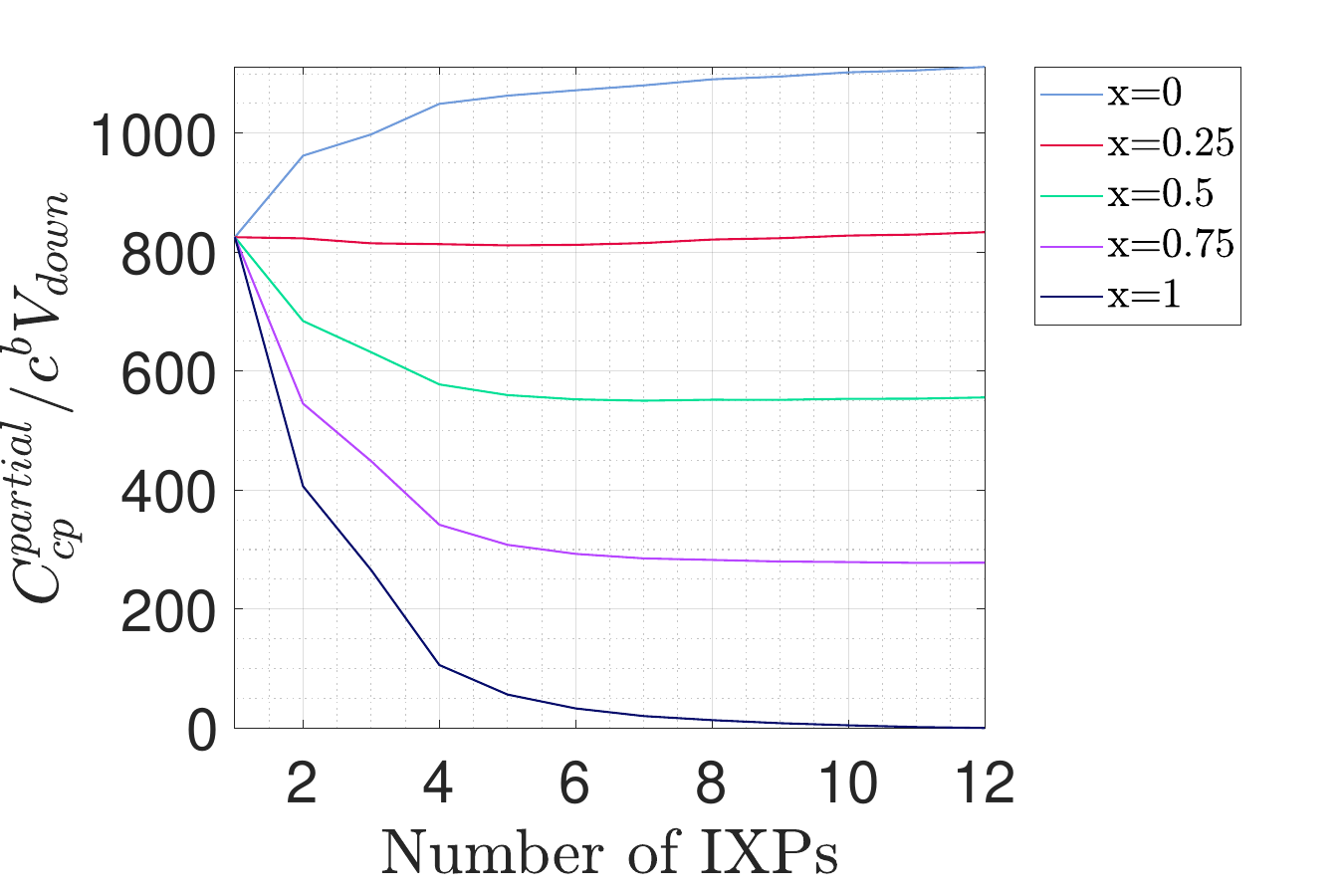}
\caption{ISP Cost}
\label{fig:CP}
\end{figure}

The effect of the number of interconnection points at which they agree to peer ($N$) on the ISP's downstream cost is illustrated in figure \ref{fig:CP}, for various values of the proportion $x$.

When $x=0$, the cost is minimized when $N=1$. We conclude an ISP has little incentive to peer at multiple IXPs with a content provider that does not replicate content and that uses hot potato routing. This is not surprising, since as we discussed in \cite{globe}, it is not rational for an ISP to agree to settlement-free peering with another ISP when the traffic ratio of downstream to upstream traffic is high. 

When $x=1$, the cost decreases as the number of IXPs increases. Note that the ISP has an incentive to increase the number of IXPs at which the two parties peer \textit{despite} the fact that the traffic ratio is infinity. However, there is less than a $2\%$ difference in the cost between $N=9$ and $N=12$, so this indicates there may be little value in requiring interconnection at more than 9 IXPs. This comparison indicates that it is likely rational for an ISP to agree to settlement-free peering with a content provider that replicates its content at all agreed peering points and delivers all traffic locally, as long as it agrees to interconnect at a minimum of 9 IXPs. We would thus expect large ISPs to have \textit{different} settlement-free peering requirements for such content providers than for ISPs. First, we would expect the minimum number of interconnection points to be higher for content providers than ISPs. Second, we would certainly expect there to be \textit{no} traffic ratio requirements for content providers. Third, we expect there to be some type of traffic localization requirement.

When $x < 0.3$, too little of the downstream traffic from the content provider to the ISP is delivered locally, and as the cost to the  ISP increases as the number of IXP increases. However, when $x > 0.3$, the ISP benefits from increasing the number of IXPs at which the two parties agree to peer. We conclude that it is likely rational for an ISP to agree to settlement-free peering with a content provider that provides partial replication and delivers that portion locally. We expect that the ISP may require a specified minimum amount of traffic to be delivered locally. We expect the ISP to require interconnection at a specified minimum number of interconnection points, although the number may depend on the amount of traffic delivered locally. However, we certainly expect there to be \textit{no} traffic ratio requirements.

\section{Conclusion}\label{sec:conclusion}

ISPs and content providers disagree about the effect of paid peering on broadband prices. ISPs assert that the revenue they generate from paid peering fees is used to lower broadband prices, whereas content providers assert that paid peering fees increase ISP profit but do not affect broadband prices. 

To address this debate, we modeled a monopoly ISP offering two tiers of service. Consumers decide whether to subscribe to broadband and if so to which tier, and whether to subscribe to video streaming services. We considered a two-sided model in which a profit-maximizing ISP determines broadband prices and the peering price and in which video streaming providers choose their price based on the peering price. Numerical parameters were chosen based on public information about broadband and video streaming prices and subscription. 

Our results show that the claims of the ISPs and of the content providers are both incorrect. When an ISP chooses peering prices, some of the revenue from these fees is used to decrease the price of the premium tier, but some of the revenue increases ISP profit. In contrast, when a regulator sets peering prices to maximize consumer surplus, the lower price stimulates significant additional demand for video streaming.

ISPs and content providers also disagree about the effect of paid peering on consumer surplus, and ultimately about whether peering prices should be regulated. Our results show that the peering price that maximizes consumer surplus is lower than the peering price an ISP would choose. Although an ISP-chosen peering price does eliminate an inherent subsidy of video streaming (if there is a positive incremental ISP cost per video streaming subscriber), the ISP-chosen peering price substantially exceeds this incremental cost. As a result, the ISP-chosen peering price reduces consumer surplus, largely because it reduces demand for video streaming. 

However, it does not follow that settlement-free peering is always the policy that maximizes consumer surplus. When there is a moderate incremental ISP cost per video streaming subscriber, the peering price that maximizes consumer surplus is positive, but lower than the ISP-chosen price. This positive price is beneficial for consumers because the incremental ISP cost for video streaming is paid by video streaming subscribers. In contrast, if content providers bring the content closer to consumers, there may be a negative incremental ISP cost per video streaming subscriber, in which case the peering price that maximizes consumer surplus is negative. In this situation, the content provider should be entitled to settlement-free peering, or even to be paid by the ISP.

In order to explain incremental ISP cost per video streaming subscriber, we examined the effect of the number of interconnection points at which a content provider peers with an ISP and video traffic localization on an ISP's variable traffic-sensitive costs. Large ISPs often argue that large content providers should meet the same requirements as other ISPs to qualify for settlement-free peering. If a content provider does not replicate its content and uses hot potato routing, an ISP is unlikely to perceive sufficient value to offer settlement-free peering. However, if a content provider does replicate its content, it is rational for an ISP to agree to settlement-free peering if the content provider agrees to interconnect at a specified minimum of IXPs and to deliver a specified minimum proportion of traffic locally.

\bibliography{References}

\begin{thebibliography}{10}
\providecommand{\url}[1]{#1}
\csname url@samestyle\endcsname
\providecommand{\newblock}{\relax}
\providecommand{\bibinfo}[2]{#2}
\providecommand{\BIBentrySTDinterwordspacing}{\spaceskip=0pt\relax}
\providecommand{\BIBentryALTinterwordstretchfactor}{4}
\providecommand{\BIBentryALTinterwordspacing}{\spaceskip=\fontdimen2\font plus
\BIBentryALTinterwordstretchfactor\fontdimen3\font minus \fontdimen4\font\relax}
\providecommand{\BIBforeignlanguage}[2]{{%
\expandafter\ifx\csname l@#1\endcsname\relax
\typeout{** WARNING: IEEEtran.bst: No hyphenation pattern has been}%
\typeout{** loaded for the language `#1'. Using the pattern for}%
\typeout{** the default language instead.}%
\else
\language=\csname l@#1\endcsname
\fi
#2}}
\providecommand{\BIBdecl}{\relax}
\BIBdecl

\bibitem{FCC}
{Federal Communications Commission}, ``\capitalisewords{Protecting and Promoting the Open Internet, Report and Order on Remand, Declaratory Ruling, and Order, 30 FCC Rcd 5601},'' 2015.

\bibitem{FCC2}
------, ``\capitalisewords{Restoring Internet Freedom, Declaratory Ruling, Report and Order, and Order, 33 FCC Rcd 311},'' 2018.

\bibitem{kim2020direct}
S.~J. Kim, ``Direct interconnection and investment incentives for content quality,'' \emph{Review of Network Economics}, vol.~18, no.~3, pp. 169--204, 2020.

\bibitem{laffont2003internet}
J.-J. Laffont, S.~Marcus, P.~Rey, and J.~Tirole, ``Internet interconnection and the off-net-cost pricing principle,'' \emph{RAND Journal of Economics}, pp. 370--390, 2003.

\bibitem{wang2018paid}
X.~Wang, Y.~Xu, and R.~T. Ma, ``Paid peering, settlement-free peering, or both?'' in \emph{IEEE INFOCOM 2018-IEEE Conference on Computer Communications}.\hskip 1em plus 0.5em minus 0.4em\relax IEEE, 2018, pp. 2564--2572.

\bibitem{musacchio2007network}
J.~Musacchio, J.~Walrand, and G.~Schwartz, ``Network neutrality and provider investment incentives,'' in \emph{2007 Conference Record of the Forty-First Asilomar Conference on Signals, Systems and Computers}.\hskip 1em plus 0.5em minus 0.4em\relax IEEE, 2007, pp. 1437--1444.

\bibitem{weisman2010price}
D.~L. Weisman and R.~B. Kulick, ``Price discrimination, two-sided markets, and net neutrality regulation,'' \emph{Tul. J. Tech. \& Intell. Prop.}, vol.~13, p.~81, 2010.

\bibitem{economides2012network}
N.~Economides and J.~T{\aa}g, ``Network neutrality on the internet: A two-sided market analysis,'' \emph{Information Economics and Policy}, vol.~24, no.~2, pp. 91--104, 2012.

\bibitem{njoroge2014investment}
P.~Njoroge, A.~Ozdaglar, N.~E. Stier-Moses, and G.~Y. Weintraub, ``Investment in two-sided markets and the net neutrality debate,'' \emph{Review of Network Economics}, vol.~12, no.~4, pp. 355--402, 2014.

\bibitem{tang2019regulating}
J.~Tang and R.~T. Ma, ``Regulating monopolistic isps without neutrality,'' \emph{IEEE Journal on Selected Areas in Communications}, vol.~37, no.~7, pp. 1666--1680, 2019.

\bibitem{ma2008interconnecting}
R.~T. Ma, D.-m. Chiu, J.~C. Lui, V.~Misra, and D.~Rubenstein, ``Interconnecting eyeballs to content: A shapley value perspective on isp peering and settlement,'' in \emph{Proceedings of the 3rd international workshop on Economics of networked systems}, 2008, pp. 61--66.

\bibitem{wu2011revenue}
Y.~Wu, H.~Kim, P.~H. Hande, M.~Chiang, and D.~H. Tsang, ``Revenue sharing among isps in two-sided markets,'' in \emph{2011 Proceedings IEEE INFOCOM}.\hskip 1em plus 0.5em minus 0.4em\relax IEEE, 2011, pp. 596--600.

\bibitem{wang2017optimal}
X.~Wang, R.~T. Ma, and Y.~Xu, ``On optimal two-sided pricing of congested networks,'' \emph{Proceedings of the ACM on Measurement and Analysis of Computing Systems}, vol.~1, no.~1, pp. 1--28, 2017.

\bibitem{chang2006peer}
H.~Chang, S.~Jamin, and W.~Willinger, ``To peer or not to peer: Modeling the evolution of the internet's as-level topology,'' in \emph{Proceedings IEEE INFOCOM}, 2006, pp. 1--12.

\bibitem{sirbu2011economic}
M.~A. Sirbu and P.~Agyapong, ``Economic incentives in content-centric networking: Implications for protocol design and public policy.''\hskip 1em plus 0.5em minus 0.4em\relax The Research Conference on Communications, Information, and Internet Policy (TPRC), 2011.

\bibitem{VerizonReplyComments}
``{Reply Comments of Verizon and Verizon Wireless, In the Matter of Framework for Broadband Internet Service (GN Docket No. 10-127) and Open Internet Rulemaking (GN Docket No. 14-28)},'' September 15, 2014.

\bibitem{NetflixComments}
``{Comments of Netflix, In the Matter of Framework for Broadband Internet Service (GN Docket No. 10-127) and Open Internet Rulemaking (GN Docket No. 14-28)},'' July 15, 2014.

\bibitem{Pew}
{Pew Research Center}, ``\capitalisewords{Internet/Broadband Fact Sheet},'' \emph{Retrieved from https://www.pewresearch.org/internet/fact-sheet/internet-broadband/. Accessed July 17, 2021}, 2021.

\bibitem{WSJ}
{The Wall Street Journal}, ``\capitalisewords{Do You Pay Too Much for Internet Service? See How Your Bill Compares},'' \emph{Retrieved from https://www.wsj.com/articles/do-you-pay-too-much-for-internet-service-see-how-your-bill-compares-11577199600. Accessed July 17, 2021}, 2019.

\bibitem{Deloitte}
{Deloitte's TMT Center}, ``\capitalisewords{Digital media trends: A look beyond generations},'' \emph{Retrieved from https://www2.deloitte.com/us/en/insights/industry/technology/digital-media-trends-consumption-habits-survey/video-streaming-wars-redrawing-battle-lines.html. Accessed July 17, 2021}, 2019.

\bibitem{Leichtman}
{Leichtman Research Group}, ``\capitalisewords{78 \% of U.S. Households Have an {SVOD} Service},'' \emph{Retrieved from https://www.leichtmanresearch.com/78-of-u-s-households-have-an-svod-service/. Accessed July 17, 2021}, 2020.

\bibitem{Parks}
{Parks Associates}, ``\capitalisewords{61 \% of US broadband households subscribe to two or more OTT services},'' \emph{Retrieved from https://www.parksassociates.com/blog/article/pr-11232020. Accessed July 17, 2021}, 2020.

\bibitem{Netflix}
Netflix, ``\capitalisewords{Plans and Pricing},'' \emph{Retrieved from https://www.netflix.com/. Accessed Oct 28, 2021}, 2021.

\bibitem{Hulu}
Hulu, ``\capitalisewords{Plans and Pricing},'' \emph{Retrieved from http://www.hulu.com/. Accessed Oct 28, 2021}, 2021.

\bibitem{Disney}
Disney+, ``\capitalisewords{Plans and Pricing},'' \emph{Retrieved from https://www.disneyplus.com/home. Accessed Oct 28, 2021}, 2021.

\bibitem{HBO}
{HBO Max}, ``\capitalisewords{Plans and Pricing},'' \emph{Retrieved from https://www.hbomax.com/. Accessed Oct 28, 2021}, 2021.

\bibitem{IXP4}
\BIBentryALTinterwordspacing
{Network Startup Resource Center (NSRC)}, ``\capitalisewords{Internet eXchange Points: North America region},'' Accessed January 26, 2022. [Online]. Available: \url{https://nsrc.org/ixp/NorthAmerica.html}
\BIBentrySTDinterwordspacing

\bibitem{IXP5}
\BIBentryALTinterwordspacing
{PeeringDB}, ``\capitalisewords{Exchanges},'' Accessed January 26, 2022. [Online]. Available: \url{https://www.peeringdb.com/advanced_search?country__in=US&reftag=ix}
\BIBentrySTDinterwordspacing

\bibitem{IXP6}
\BIBentryALTinterwordspacing
{Hurricane Electric {Internet} Services}, ``\capitalisewords{Internet Exchange Report},'' Accessed January 26, 2022. [Online]. Available: \url{https://bgp.he.net/report/exchanges}
\BIBentrySTDinterwordspacing

\bibitem{Lat_Long}
\BIBentryALTinterwordspacing
{United States Census Bureau}, ``{The U.S. Gazetteer Files},'' Accessed January 26, 2022. [Online]. Available: \url{https://www.census.gov/geographies/reference-files/time-series/geo/gazetteer-files.2021.html}
\BIBentrySTDinterwordspacing

\bibitem{Census}
\BIBentryALTinterwordspacing
------, ``{Vintage county population estimates totals},'' Accessed January 26, 2022. [Online]. Available: \url{https://www.census.gov/programs-surveys/popest/technical-documentation/research/evaluation-estimates/2020-evaluation-estimates/2010s-counties-total.html}
\BIBentrySTDinterwordspacing

\bibitem{valancius2011many}
V.~Valancius, C.~Lumezanu, N.~Feamster, R.~Johari, and V.~V. Vazirani, ``How many tiers? pricing in the {Internet} transit market,'' in \emph{Proceedings of the ACM Special Interest Group on Data Communication Conference (SIGCOMM)}, 2011, pp. 194--205.

\bibitem{globe}
A.~Nikkhah and S.~Jordan, ``Requirements of settlement-free peering policies,'' in \emph{2022 IEEE Global Communications Conference (GLOBECOM)}, 2022.

\end{thebibliography}
\bibliographystyle{IEEEtran}

\end{document}